\documentclass[twocolumn,english,reprint, superscriptaddress, breaklinks=true, showkeys, showpacs=false, nofootinbib]{revtex4}

\usepackage[T1]{fontenc}
\usepackage{color}
\usepackage{babel}
\usepackage{float}
\usepackage{amssymb}
\usepackage{listings}
\usepackage{graphicx}
\usepackage{subcaption}
\usepackage{amsmath}
\usepackage{amsthm}
\usepackage{svg}
\usepackage{caption}
\usepackage{mathtools}
\usepackage{ragged2e}
\usepackage{soul}

\usepackage[unicode=true,pdfusetitle, bookmarks=true,bookmarksnumbered=false,bookmarksopen=false, breaklinks=true,pdfborder={0 0 0},backref=false,colorlinks=true]{hyperref}
\usepackage{breakurl}

\definecolor{Code}{rgb}{0,0,0}
\definecolor{Decorators}{rgb}{0.5,0.5,0.5}
\definecolor{Numbers}{rgb}{0.5,0,0}
\definecolor{MatchingBrackets}{rgb}{0.25,0.5,0.5}
\definecolor{Keywords}{rgb}{0,0,1}
\definecolor{self}{rgb}{0,0,0}
\definecolor{Strings}{rgb}{0,0.63,0}
\definecolor{Comments}{rgb}{0,0.63,1}
\definecolor{Backquotes}{rgb}{0,0,0}
\definecolor{Classname}{rgb}{0,0,0}
\definecolor{FunctionName}{rgb}{0,0,0}
\definecolor{Operators}{rgb}{0,0,0}
\definecolor{Background}{rgb}{0.98,0.98,0.98}
\lstdefinelanguage{Python}{
numbers=left,
numberstyle=\footnotesize,
numbersep=1em,
xleftmargin=1em,
framextopmargin=2em,
framexbottommargin=2em,
showspaces=false,
showtabs=false,
showstringspaces=false,
frame=l,
tabsize=4,
basicstyle=\ttfamily\small\setstretch{1},
backgroundcolor=\color{Background},
commentstyle=\color{Comments}\slshape,
stringstyle=\color{Strings},
morecomment=[s][\color{Strings}]{"""}{"""},
morecomment=[s][\color{Strings}]{'''}{'''},
morekeywords={import,from,class,def,for,while,if,is,in,elif,else,not,and,or,print,break,continue,return,True,False,None,access,as,,del,except,exec,finally,global,import,lambda,pass,print,raise,try,assert},
keywordstyle={\color{Keywords}\bfseries},
morekeywords={[2]@invariant,pylab,numpy,np,scipy},
keywordstyle={[2]\color{Decorators}\slshape},
emph={self},
emphstyle={\color{self}\slshape},
}
\makeatletter
\@ifundefined{textcolor}{}
{%
 \definecolor{BLACK}{gray}{0}
 \definecolor{WHITE}{gray}{1}
 \definecolor{RED}{rgb}{1,0,0}
 \definecolor{GREEN}{rgb}{0,1,0}
 \definecolor{BLUE}{rgb}{0,0,1}
 \definecolor{CYAN}{cmyk}{1,0,0,0}
 \definecolor{MAGENTA}{cmyk}{0,1,0,0}
 \definecolor{YELLOW}{cmyk}{0,0,1,0}
}
\hypersetup{colorlinks=true,citecolor=blue,linkcolor=cyan,urlcolor=blue,filecolor= green, breaklinks=true}
\usepackage{url}
\usepackage{breakurl}

\makeatother

\begin{document}

\title{Prime Number Identification Demonstrated with Quantum Processors \\ Using a New Rescaling-Based Noise Mitigation Technique}

\author{Victor F. dos Santos\href{https://orcid.org/0009-0009-0319-4852}{\includegraphics[scale=0.05]{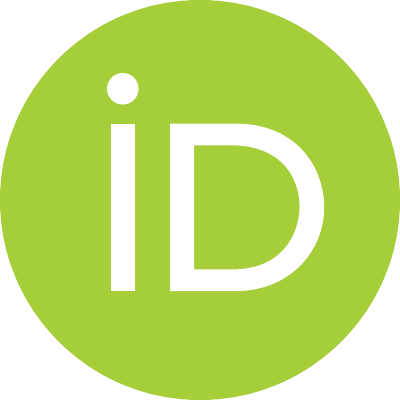}}}
\email[]{victorfds997@gmail.com}
\affiliation{Physics Department, 
Federal University of Santa Maria, 97105-900,
Santa Maria, RS, Brazil}

\author{Victor P. Brasil\href{https://orcid.org/0009-0009-8203-3976}{\includegraphics[scale=0.05]{orcidid.pdf}}}
\email[]{victor.brasil@acad.ufsm.br}
\affiliation{Physics Department, 
Federal University of Santa Maria, 97105-900,
Santa Maria, RS, Brazil}

\author{Pedro A. S. Contri\href{https://orcid.org/0009-0007-4519-9745}{\includegraphics[scale=0.05]{orcidid.pdf}}}
\email[]{pedro.contri@acad.ufsm.br}
\affiliation{Physics Department, 
Federal University of Santa Maria, 97105-900,
Santa Maria, RS, Brazil}

\author{Jonas Maziero\href{https://orcid.org/0000-0002-2872-986X}{\includegraphics[scale=0.05]{orcidid.pdf}}}
\email[]{jonas.maziero@ufsm.br}
\affiliation{Physics Department, 
Federal University of Santa Maria, 97105-900,
Santa Maria, RS, Brazil}

\begin{abstract}
We implement a quantum protocol for prime number identification based on entanglement
dynamics, using IBM quantum processors. The method links the primality of an integer to specific Fourier components extracted from the time evolution of entanglement in a bipartite quantum system. To mitigate experimental noise,
we introduce a noise-mitigation method based on a global rescaling factor, which is
calibrated on a subset of circuits and extrapolated across different configurations.
Theoretical support is provided by a new analytical bound for the Fourier modes derived assuming an initial uniform superposition state. This new bound enhances the separation between prime and composite numbers under moderate experimental deviations. These results represent a step toward practical number-theoretic applications on noisy intermediate-scale quantum (NISQ) devices.
\end{abstract}


\maketitle

\section{Introduction}
\label{sec:introduction}
The identification and classification of prime numbers is a fundamental problem in number theory~\cite{hardywright2008, ford2016,maynard2019,lichtman2023,wu2025}, with far-reaching implications in cryptography, computational complexity, and theoretical computer science~\cite{shor1997}. While classical algorithms for primality testing have advanced significantly~\cite{aks2004}, quantum approaches offer alternative perspectives by encoding number-theoretic properties into quantum observables~\cite{latorre2014,miller2019}. In particular, entanglement-based quantum protocols have shown promise in revealing arithmetic structure through measurable quantum correlations~\cite{latorre2015,duchi2020}. Connections between the distribution of primes and physical spectral phenomena have also been explored from a quantum-mechanical perspective \cite{schumayer2011,creffield2015}.

In Ref.~\cite{santos2024}, building on the earlier work of Ref.~\cite{southier2023}, two of us introduced a quantum algorithm, named here as PIED (Prime Identification via Entanglement Dynamics), that associates the prime nature of integers with signatures in the Fourier spectrum of the reduced purity (the quantum entanglement) of a bipartite quantum system~\cite{zylberman2024}. 
Unlike Shor's algorithm, which exploits algebraic periodicity through quantum Fourier transforms for factoring and discrete logarithms, PIED does not implement a period-finding routine. Rather, it follows an observable-based approach in which arithmetic information is encoded in entanglement dynamics and accessed through the spectral components of a measurable quantity. This places the method closer in spirit to quantum number-theoretic frameworks based on prime states and spectral signatures of arithmetic structure, while also suggesting connections with spectral-estimation techniques.
The algorithm begins with an initial product state evolved under a diagonal Hamiltonian with uniformly spaced energy levels~\cite{nielsenchuang2010}. Although the simulations in that study focused on the uniform superposition state, the formulation is general and permits arbitrary product state initialization. The information about an integer \(n\) is then extracted by analyzing the amplitudes of specific frequency components that arise in the resulting entanglement dynamics. 
Previous work showed that prime numbers correspond to minimal values of the Fourier
modes, while composite numbers yield systematically larger values.

In this work, we implement the algorithm on IBM Quantum processors, operating within the noisy intermediate-scale quantum (NISQ) regime~\cite{preskill2018}, thus transitioning from classical simulation to real hardware~\cite{russo2024}. To mitigate the effects of noise in the measurement of the reduced purity, we develop and explore a specific approach, termed CFE (Correction Factor Extrapolation), which involves a global rescaling applied to the Fourier modes obtained from the purity function. This method requires only a single set of circuit executions for each chosen value of a given parameter and leverages the resulting Fourier-mode deviations to construct a corresponding correction factor, \(\Lambda_{\text{opt}}(\lambda)\), that can be extrapolated across different parameter values. As such, it offers a computationally inexpensive alternative to traditional noise-mitigation strategies in algorithms where the same physical quantity can be obtained for equivalent quantum circuits with varying parameters, such as, in the present case, the system dimension.

We also present a refined theoretical result, as detailed in Appendix \ref{appendix:composite_bound}, which strengthens the separation between prime and composite integers by introducing a second analytical lower bound for the Fourier modes \(\alpha_n\), proven for the case where the initial state is the uniform superposition state. This bound corresponds to the \(\alpha_n\) for integers of the form \(n = k^2\), where \(k\) is prime, and holds for the majority of composite numbers. We show that only composite integers equal to either a semiprime of 2 or 3 can produce values of \(\alpha_n\) that fall below the second bound. The corresponding Fourier-mode curves for these families are analytically characterized, and we obtain the threshold value \(n_{\text{th}}\) beyond which they begin to undercut the bound. This classification significantly narrows the set of exceptions, providing a clearer interpretive framework: when an experimentally measured mode \(\alpha_n\) falls below or close to the second bound, it suffices to check whether \(n\) belongs to one of the three known types (semiprime of 2, semiprime of 3, or new-bound case). If not, the value is very likely associated with a prime number. Although the formal proof applies only to the uniform superposition state, our observations suggest that a broader class of initial states, possibly characterized by specific amplitude distributions, may exhibit similar behavior, an open direction for future analysis.

The remainder of this article is organized as follows. In Sec.~\ref{sec:theoretical}, we briefly summarize the theoretical framework and review the main elements of PIED (Prime Identification via Entanglement Dynamics) algorithm, including a discussion on the new analytical bound for the Fourier modes that enhances the prime/composite separation. The proof for this bound is detailed in Appendix \ref{appendix:composite_bound}. Section~\ref{sec:results} presents our implementation of PIED on IBM Quantum hardware and evaluates the proposed noise-mitigation strategy, CFE (Correction Factor Extrapolation), using deviations in Fourier modes as a primary diagnostic, though the method is not limited to this quantity. Complementary results using zero-noise extrapolation are reported in Appendix~\ref{appendix:zne}. We conclude in Sec.~\ref{sec:conclusions} with a discussion of the implications of our findings and possible future directions. Additional results for spin coherent initial states are presented in Appendix~\ref{appendix:spin}.

\section{Algorithm Summary and Theoretical Background}
\label{sec:theoretical}
We begin by reviewing the PIED algorithm, which is built on two key components: a bipartite quantum system \( AB \), composed of two identical subsystems \( A \) and \( B \), each of dimension \( d \); and a time-independent Hamiltonian defined as \begin{equation} 
\hat{H}_{AB} = g \bigl(\hat{H}_A \otimes \hat{H}_B\bigr), \end{equation}
where \( g \) is a coupling constant, and \( \hat{H}_A = \hat{H}_B \) is a local Hamiltonian with eigenvalues of the form \( n_S \mu \), where \( n_S \in \{1, 2, \dots, d\};\ S \in \{A, B\};\) and \( \mu \) is a fixed energy spacing. When implementing the algorithm on qubit-based quantum hardware, we take the subsystem dimension \(d\) to be a power of two, so that the bipartite Hilbert space is naturally represented on a register of qubits. This choice ensures that all computational basis states are consistently included in the state preparation, time evolution, and purity evaluation. Target values of \(d\) that are not powers of two can be addressed by implementing the algorithm at the nearest admissible dimension compatible with qubit registers.

The initial state is taken to be a product state of the form \( |\psi(0)\rangle_{AB} = |\phi\rangle_A \otimes |\phi\rangle_B \), where \( |\phi\rangle \) is a pure state in a \( d \)-dimensional Hilbert space. Using the standard computational basis \( \{ |E_{n_S}\rangle \} \), which also coincides with the eigenbasis of the local Hamiltonian \( \hat{H}_{S} \), we encode each \( |\phi\rangle_S = \sum_{n_S =1}^d c_{n_S}|E_{n_S}\rangle \) into a multi-qubit register via a fixed mapping from dimension \( d \) to \( \lceil \log_2 d \rceil \) qubits. The probability amplitudes \( c_{n_S} \) associated with each basis state \( |E_{n_S}\rangle \) must satisfy the condition \( c_{n_S} \neq 0 \) for all \( n_S \). This non-vanishing condition ensures that all energy levels contribute to the subsequent system dynamics. From these conditions, we adopt the simplest choice of initial state: the uniform superposition state. In this case, the coefficients satisfy \( c_{n_S} = \frac{1}{\sqrt{d}} \) for every \( n_S \), and the state can be prepared by applying a Hadamard gate to each qubit initialized in the ground state \( |0\rangle \).

Combining the system definitions and initial conditions, we obtain the unitary operator that governs the time evolution of the composite system:
\begin{align}
    \hat{U}(t)_{AB} &= e^{-i \hat{H}_{AB} t / \hbar} \notag \\
    &= \sum_{n_A, n_B = 1}^{d} e^{-i \omega n_A n_B t} \, |E_{n_A}E_{n_B}\rangle \langle E_{n_A}E_{n_B}|,
\end{align}
where we have defined \( \omega = g \mu^2 / \hbar \). As shown in Ref.~\cite{santos2024}, this unitary operator can be efficiently implemented using the protocol presented in Ref.~\cite{welch2014}. To achieve this, we employ the formalism of Walsh functions~\cite{perez2022}, which enables the implementation of any diagonal unitary gate using only $Z$ rotations and $CNOT$ gates~\cite{shende2006}, with a circuit depth that scales polynomially in the number of qubits in this case~\cite{zhang2024,huang2024jpsj}.

From the evolution operator \( \hat{U}(t)_{AB} \), we obtain the state of the system at time \( t \):
\begin{align}
|\psi(t)\rangle_{AB} &= \hat{U}(t)_{AB} |\psi(0)\rangle \notag \\ 
&= \sum_{n_A, n_B = 1}^{d} c_{n_A} c_{n_B} e^{-i \omega n_A n_B t} \, |E_{n_A} E_{n_B}\rangle.
\end{align}
This expression allows us to access information about subsystem \( A \) by computing its reduced density matrix, defined as
\begin{equation}
\hat{\rho}_{A}(t) = \operatorname{Tr}_B \left( |\psi(t)\rangle_{AB} \langle \psi(t)| \right).
\end{equation}
The purity of subsystem \( A \), which quantifies the degree of entanglement between \( A \) and \( B \), is then given by
\begin{align}
\gamma_A(t) &= \operatorname{Tr} \left( \hat{\rho}^2_A(t) \right) \notag \\
&= \sum_{j, k, l, m = 1}^d |c_j|^2 |c_k|^2 |c_l|^2 |c_m|^2 \, e^{-i \omega t (j - k)(l - m)}.
\end{align}
In this expression, the phase factors $e^{-i\omega t (j-k)(l-m)}$ occur in matched index pairs, such that for every term with phase $+\phi$ there is a corresponding term with phase $-\phi$. These pairs contribute equal and opposite imaginary parts, which cancel exactly for each fixed value of $t$. Therefore $\gamma_A(t)$ is strictly real for all times, and the Fourier transform of $\gamma_A(t)$ is also purely real. No information is lost by restricting the analysis to the real part of the Fourier spectrum.

An efficient way to measure \( \gamma_A(t) \) is to implement the SWAP test protocol~\cite{ekert2002}, which requires preparing two identical copies of the system \( AB \), each initialized in the same state and evolved under the same unitary operator. A single ancilla qubit is then used to control a sequence of SWAP operations between the two subsystems, ultimately encoding the value of \( \gamma_A(t) \) in the probability \( P_0(t) \) of measuring the ancilla in the state \( |0\rangle \):
\begin{equation}
\gamma_A(t) = 2P_0(t) - 1.
\end{equation}
Although alternative approaches exist for estimating the purity from single-copy measurements---such as randomized and classical-shadow protocols~\cite{elben2018,brydges2019,huang2020,elben2023,tan2021}---in this work we employ the SWAP test, since it provides a direct and resource-efficient implementation for our algorithm. In particular, the SWAP test minimizes circuit-depth growth while requiring measurements only on the ancilla qubit, from which the desired probability is statistically estimated. This makes it especially suitable for implementations on current quantum hardware. Once the purity is obtained from these implementations, its behavior can be analyzed as a function of time. As a consequence of the structure of \( \hat{H}_{AB} \), the purity \( \gamma_A(t) \) is a real and periodic function with period \( T = 2\pi/\omega \), and is symmetric around \( T/2 \).

To extract number-theoretic information from the purity dynamics, we calculate the Fourier modes \( \alpha_n \) of \( \gamma_A(t) \) using
\begin{align}
\alpha_n &= \frac{2\omega}{\pi} \int_0^{\pi/\omega} \gamma_A(t) \cos(n\omega t) \, dt \notag \\
&= 4\sum_{s = 1}^{z}\sum_{k = 1}^{d - \frac{n}{y^{(n)}_s}}\sum_{m = 1}^{d - y^{(n)}_s}|c_{k}|^2|c_{m}|^2|c_{k + \frac{n}{y^{(n)}_s}}|^2|c_{m + y^{(n)}_s}|^2,
\end{align}
where \( z \) is the number of distinct positive divisors of \( n \), and \( y^{(n)}_s \) denotes the \( s \)-th divisor, ordered increasingly so that \( y^{(n)}_1 = 1 \) and \( y^{(n)}_z = n \). The Fourier modes \( \alpha_n \) admit a minimal value \( B_n \), which corresponds to the contribution from the trivial divisors of \( n \), where \( z = 2 \). This bound is realized precisely when \( n \) is a prime number and is given by
\begin{equation}
B_n = 8 \sum_{k = 1}^{d - n} \sum_{m = 1}^{d - 1} |c_k|^2 |c_m|^2 |c_{k + n}|^2 |c_{m + 1}|^2,
\end{equation}
valid for \( 2 \leq n \leq d - 1 \). For \( n \geq d \), we define \( B_n = 0 \), since the product \( (j - k)(l - m) \) cannot reach such values within the index ranges allowed by the purity expression. Thus, \( \alpha_n \) can also be written as
\begin{equation}
\alpha_n = B_n + 4 \sum_{s = 2}^{z - 1} \sum_{k = 1}^{d - \frac{n}{y^{(n)}_s}} \sum_{m = 1}^{d - y^{(n)}_s}
    |c_k|^2 |c_m|^2 |c_{k + \frac{n}{y^{(n)}_s}}|^2 |c_{m + y^{(n)}_s}|^2.
\end{equation}
This decomposition isolates the contribution from composite structure and makes it explicit that
\begin{equation}
    \alpha_n \geq B_n,
\end{equation}
with equality if and only if \( n \in [2, 2(d-1)] \) is a prime number. This establishes \( B_n \) as not only a theoretical lower bound for the Fourier modes, but also their exact value in the case of primes, underscoring its key role in the algorithm’s identification of primality. 

\begin{figure*}
    \centering
    \includegraphics[width=0.9\textwidth]{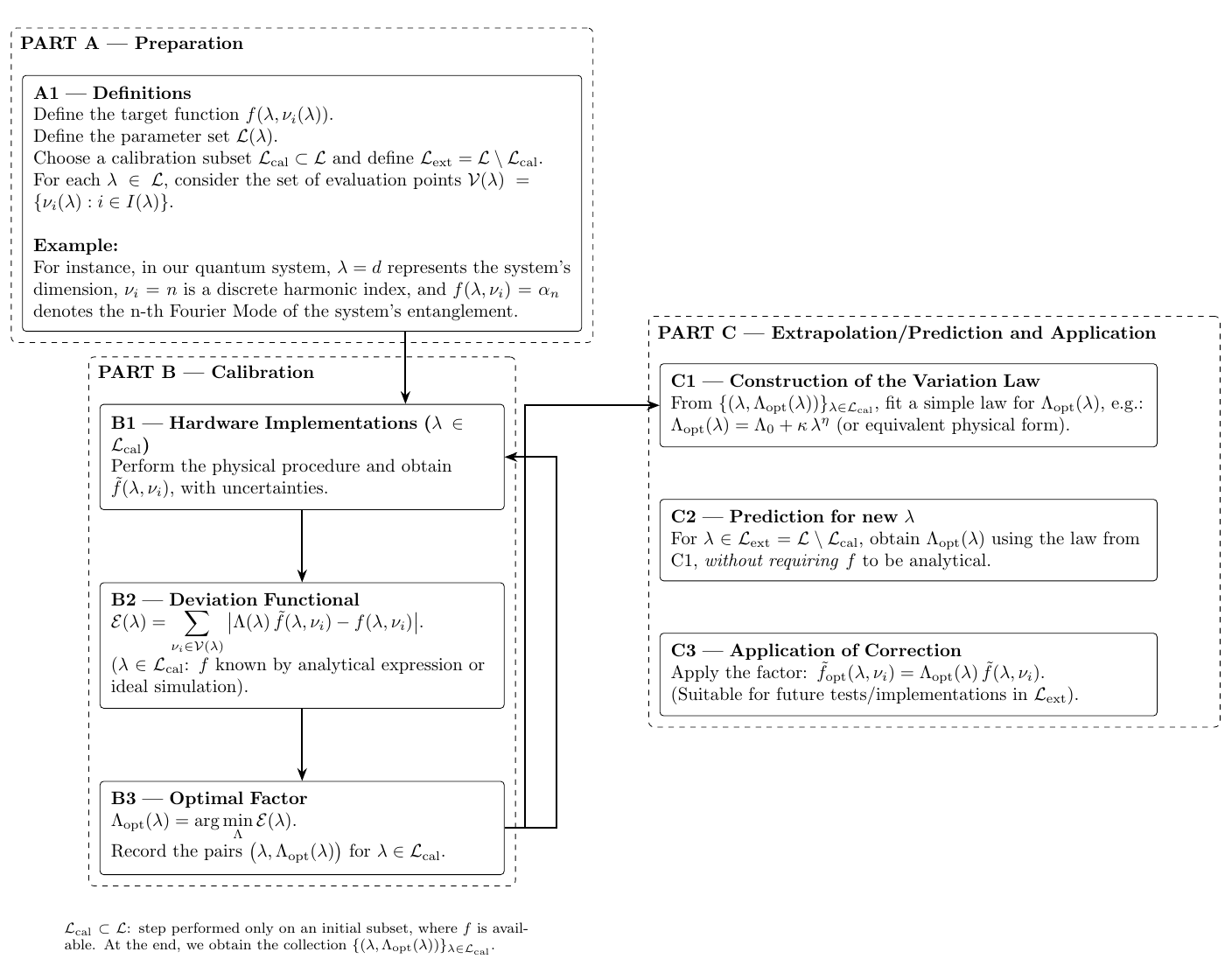}

    \vspace{1ex}
    \refstepcounter{figure}%
    \begin{minipage}[t]{0.9\textwidth}
        \justifying
        FIG.~\thefigure. Schematic representation of the Correction Factor Extrapolation (CFE) procedure. The method is divided into three stages. In part A (Preparation), the target function \(f(\lambda,\nu_i)\), the parameter set \(\mathcal{L}\), and the calibration subset \(\mathcal{L}_{\mathrm{cal}}\) are defined. In this work we set \(f(\lambda,\nu_i)=\alpha_{n}\), i.e., the Fourier-mode amplitude \(\alpha_n\) of the purity curve; \(\lambda=d\) is the system dimension and \(\nu_i=n\) is the Fourier-mode index. In part B (Calibration), experimental data \(\tilde f(\lambda,\nu_i)\) are obtained for \(\lambda \in \mathcal{L}_{\mathrm{cal}}\), and the optimal correction factors \(\Lambda_{\mathrm{opt}}(\lambda)\) are determined by minimizing the deviation function \(\mathcal{E}(\lambda)\). In part C (Extrapolation and Application), the extrapolated region \(\mathcal{L}_{\mathrm{ext}} = \mathcal{L} \setminus \mathcal{L}_{\mathrm{cal}}\) is considered, where \(\Lambda_{\mathrm{opt}}(\lambda)\) is fitted with a simple law (for example, \(\Lambda_{\mathrm{opt}}(\lambda)=\Lambda_0+\kappa \lambda^{\eta}\)) and then used to rescale new measurements as \(\tilde f_{\mathrm{opt}}(\lambda,\nu_i)=\Lambda_{\mathrm{opt}}(\lambda)\,\tilde f(\lambda,\nu_i)\).
    \end{minipage}
    \label{fig:diagram_cfe}
\end{figure*}
In practical scenarios, the values of \( \alpha_n \) are computed numerically by discretizing the integral of \( \gamma_A(t) \cos(n \omega t) \) over time into \( p \) partitions. The number \( p \) controls the integration accuracy and typically grows with the system dimension \( d \). Moreover, the required number of partitions depends on the choice of initial state: smoother purity curves tend to allow accurate estimation with fewer points. While our main analysis focuses on the uniform superposition, we also investigate an alternative initialization based on spin coherent states. These exhibit empirically favorable scaling in \( p \) with respect to \( d \), although this approach is currently limited to dimensions for which explicit implementations of spin coherent states are feasible, as their preparation becomes increasingly challenging for larger \( d \).

To better distinguish prime and composite integers within the Fourier spectrum, we derived in Appendix \ref{appendix:composite_bound} an additional analytical lower bound \(P_n\), valid for composite integers. 
This bound corresponds to the Fourier-mode amplitudes \(\alpha_n\) obtained for integers of the form \(n = k^2\), where \(k\) is prime. 
Other composite integers can yield values of \(\alpha_n\) between \(B_n\) and \(P_n\), but only in specific subintervals of \(n\) corresponding to semiprimes built from the smallest primes, \(n = 2k\) and \(n = 3k\). 
In these regions, distinct divisibility tests must be performed depending on the interval in which \(n\) lies: for the first subinterval, divisibility by \(\sqrt{n}\) is tested; for the second, by \(2\) and by \(\sqrt{n}\); and for the third, by \(2\), \(3\), and \(\sqrt{n}\). 
Small experimental deviations may also lead to \(\alpha_n\) values for \(n = k^2\) slightly above \(B_n\), which can be treated under the same criteria. 
Altogether, the bound \(P_n\) characterizes the Fourier-mode amplitudes associated with \(n = k^2\) and delineates the narrow region where semiprime-related and near-ideal cases may appear, thereby refining the identification of composite numbers in the Fourier spectrum.

\section{Development and Implementation of the CFE Method}
\label{sec:results}
In what follows, we present the results obtained from our hardware implementations on IBM Quantum processors. To mitigate the impact of noise on these results, we develop and apply the CFE (Correction Factor Extrapolation) method, which is a rescaling-based error mitigation strategy that produces a parameter-dependent correction factor for a quantum-measured function by fitting to known analytical behavior and extrapolating this factor across the parameter space.

\subsection{CFE formulation}
\label{sec:mitigation}
\begin{figure*}
    \centering

    \makebox[\textwidth][c]{%
        \begin{subfigure}[t]{0.4\textwidth}
            \centering
            \includegraphics[width=\linewidth]{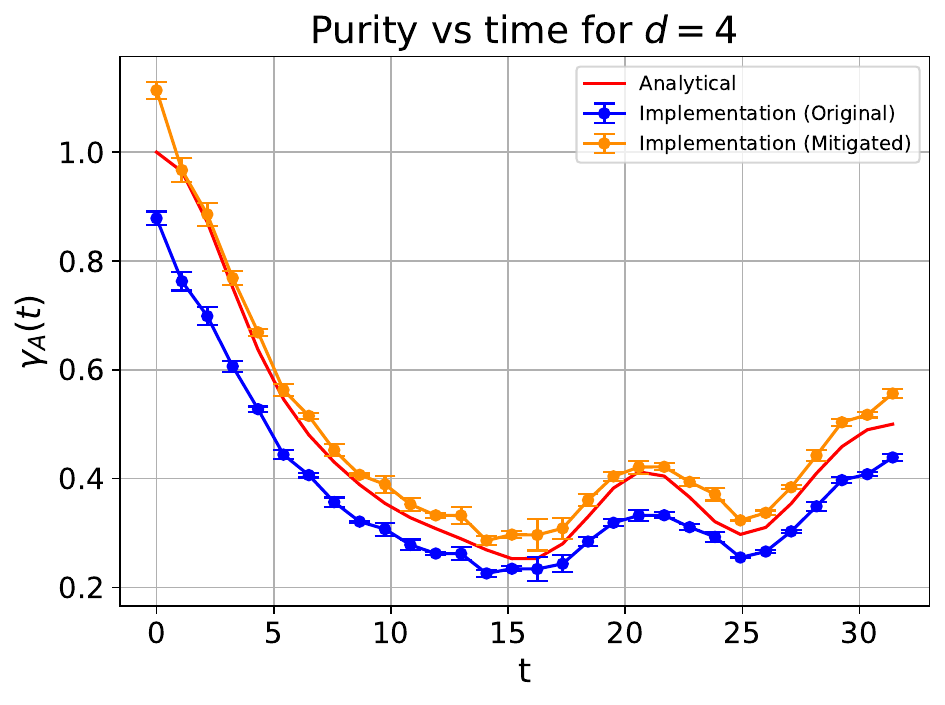}
            \caption{Time evolution of the purity of subsystem \(A\) for \( d = 4 \), obtained from quantum hardware implementations of the PIED circuit initialized with uniform superposition states.}
            \label{fig:purity_hadam_d4}
        \end{subfigure}
        \hspace{0.05\textwidth}
        \begin{subfigure}[t]{0.4\textwidth}
            \centering
            \includegraphics[width=\linewidth]{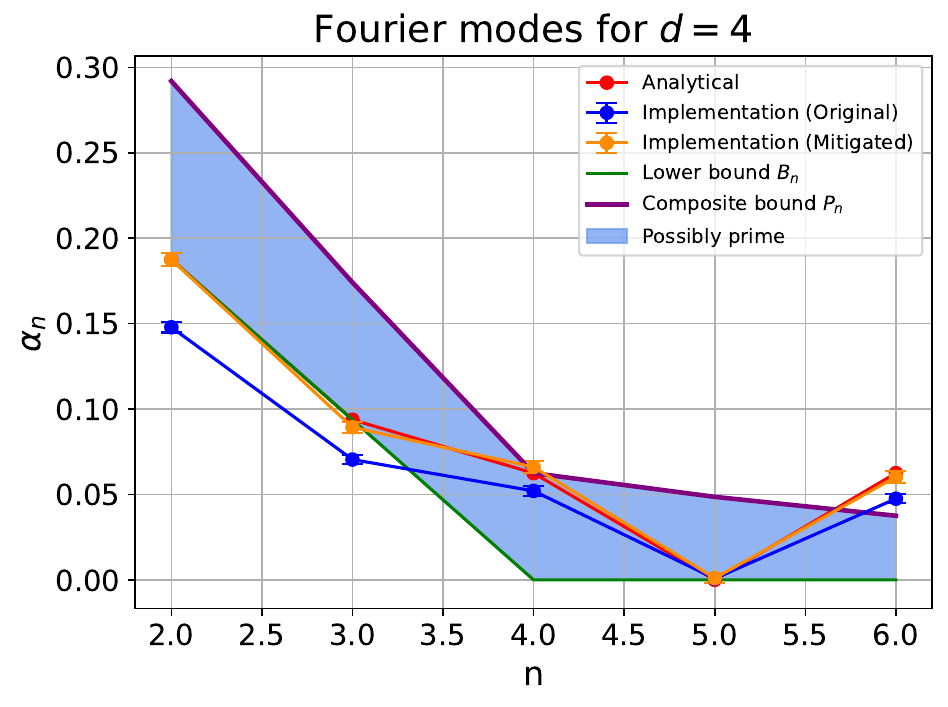}
            \caption{Fourier components extracted from the implementation  of panel~(a) for \( d = 4 \).}
            \label{fig:fourier_hadam_d4}
        \end{subfigure}
    }

    \vspace{0.4cm}

    \makebox[\textwidth][c]{%
        \begin{subfigure}[t]{0.4\textwidth}
            \centering
            \includegraphics[width=\linewidth]{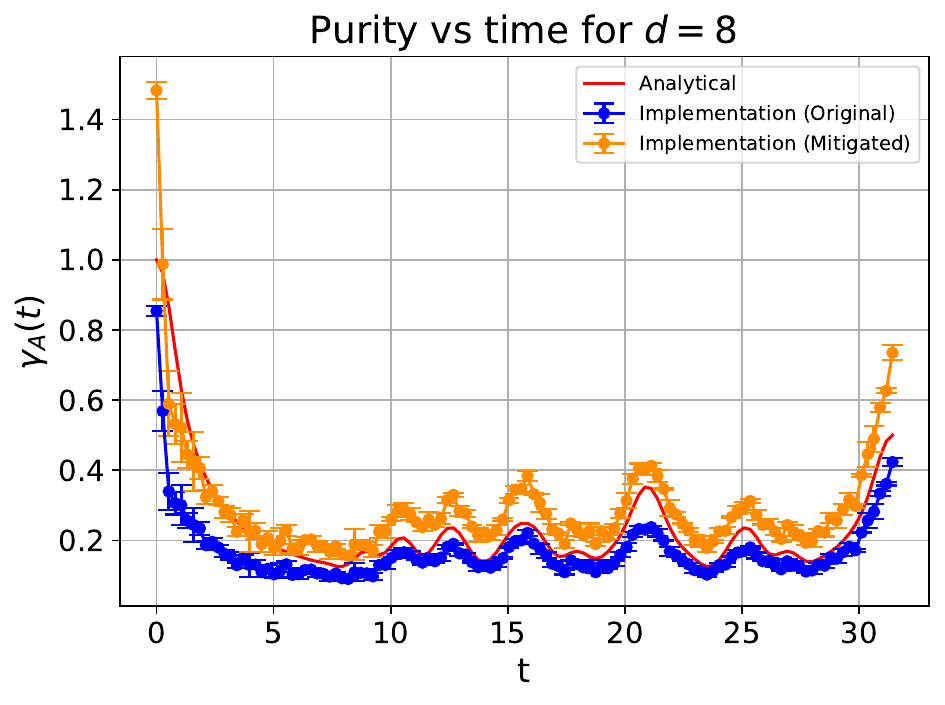}
            \caption{Time evolution of the purity of subsystem \(A\) for \( d = 8 \), obtained from quantum hardware implementations of the PIED circuit initialized with uniform superposition states.}
            \label{fig:purity_hadam_d8}
        \end{subfigure}
        \hspace{0.05\textwidth}
        \begin{subfigure}[t]{0.4\textwidth}
            \centering
            \includegraphics[width=\linewidth]{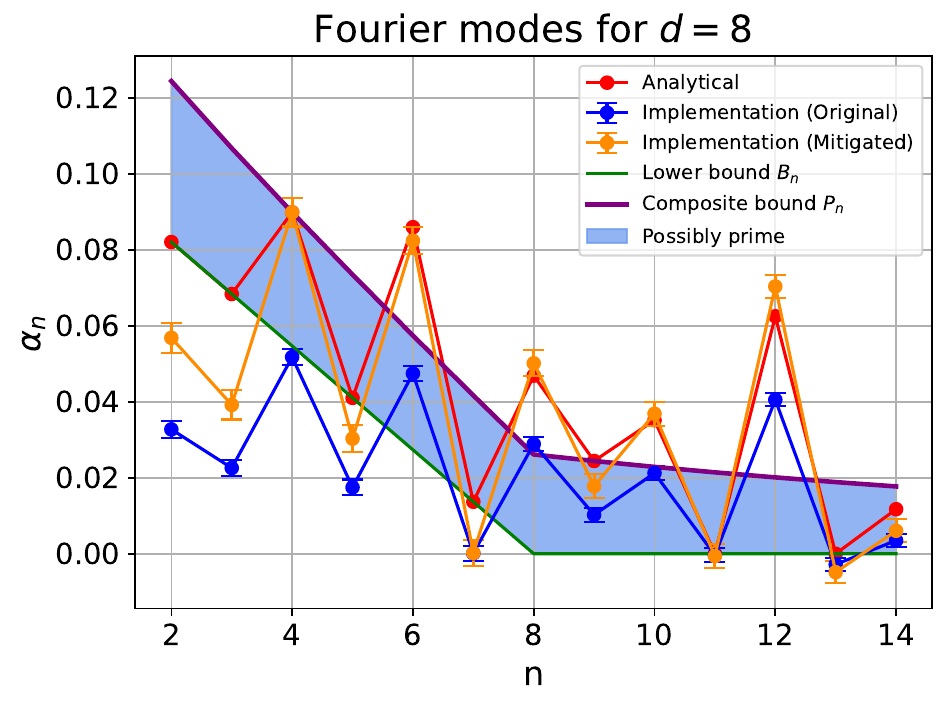}
            \caption{Fourier components extracted from the implementation  of panel~(c) for \( d = 8 \).}
            \label{fig:fourier_hadam_d8}
        \end{subfigure}
    }

    \vspace{0.4cm}

    \makebox[\textwidth][c]{%
        \begin{subfigure}[t]{0.4\textwidth}
            \centering
            \includegraphics[width=\linewidth]{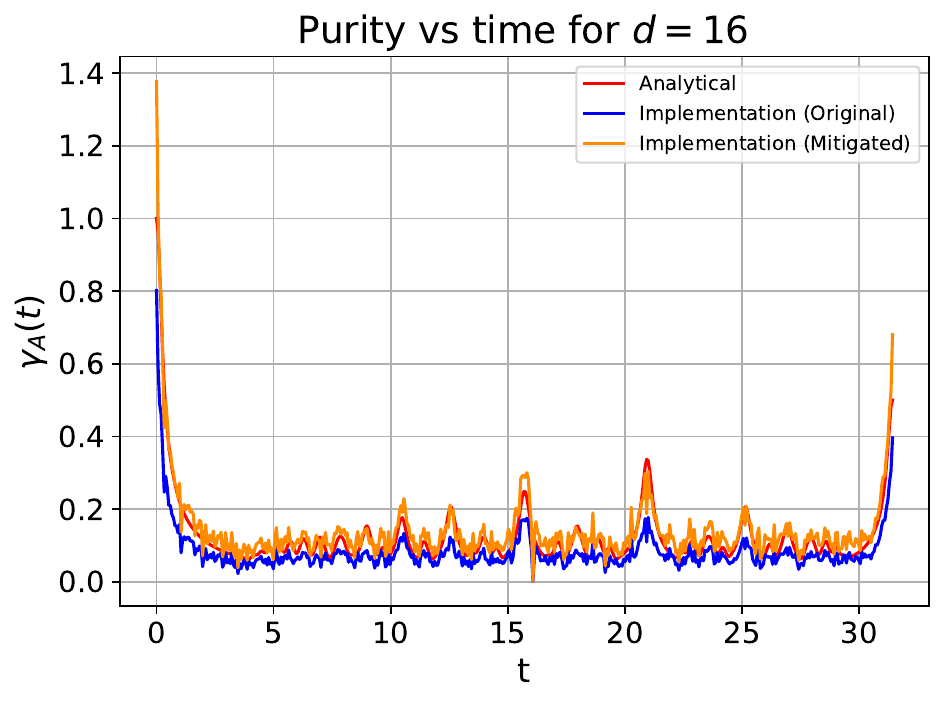}
            \caption{Time evolution of the purity of subsystem \(A\) for \( d = 16 \), obtained from quantum hardware implementations of the PIED circuit initialized with uniform superposition states.}
            \label{fig:purity_hadam_d16}
        \end{subfigure}
        \hspace{0.05\textwidth}
        \begin{subfigure}[t]{0.4\textwidth}
            \centering
            \includegraphics[width=\linewidth]{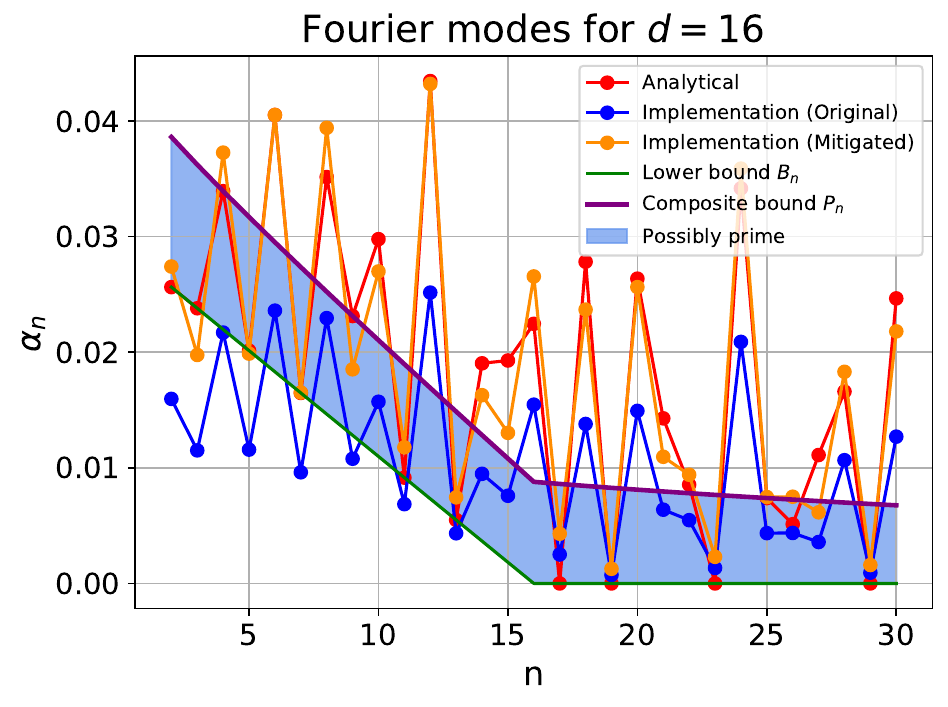}
            \caption{Fourier components extracted from the implementation  of panel~(e) for \( d = 16 \).}
            \label{fig:fourier_hadam_d16}
        \end{subfigure}
    }

    \vspace{1ex}
    \addtocounter{figure}{-1}
    \refstepcounter{figure}%
    \begin{minipage}[t]{0.9\textwidth}
        \justifying
        FIG.~\thefigure. Reduced purity (left column) and Fourier modes (right column) for \(d = 4, 8,\) and \(16\), obtained from quantum-computer implementations using \(p = 30, 120,\) and \(480\) time points, respectively, and \(2^{13}\) shots for every dimension. Each row corresponds to a different value of \(d\), illustrating how the system’s behavior changes with increasing dimensionality. In both purity and Fourier-modes plots, the red points represent the theoretical predictions from the algorithm, the blue points correspond to the results obtained directly from the hardware implementations, and the orange points show the mitigated results after applying our post-processing technique (CFE). The previously developed analysis for the positioning of integers \(n\) in the Fourier-modes plots shows strong agreement with the expected property that these positions reflect whether \(n\) is prime or composite. Simulated data are not shown, as they coincide almost perfectly with the analytical predictions.
   
    \end{minipage}   \label{fig:purity_fourier}
\end{figure*}

We now formalize the CFE procedure, schematically represented in Fig.~\ref{fig:diagram_cfe}. 
The method introduces a correction factor 
\(\Lambda_{\text{opt}}(\lambda)\) applied to noisy estimates 
\(\tilde{f}(\lambda,\nu_i(\lambda))\). 
This factor is defined through the minimization of deviations with respect to the 
corresponding analytical values \(f(\lambda, \nu_i(\lambda))\), as detailed below. 
To start, let us define a two-parameter function 
\(f(\lambda, \nu_i(\lambda))\), well behaved for any \(\lambda \in \mathcal{L}\) and 
\(\nu_i(\lambda) \in \mathcal{V}(\lambda)\), where 
\begin{equation}
\mathcal{L} := \{\lambda_1,\lambda_2,\dots,\lambda_r\},
\end{equation}
and 
\begin{equation}
\mathcal{V}(\lambda) := \{\nu_i(\lambda) : i \in I(\lambda)\},   
\end{equation} 
with \(I(\lambda)\) denoting the index set corresponding to the admissible \(\nu\)-values for each \(\lambda\).
That is, for a given \(\lambda\) in a quantum circuit, there is a finite set \(\mathcal{V}(\lambda)\) 
of associated parameters \(\nu_i(\lambda)\), typically linked to quantities extracted from measurements.
Within this general set \(\mathcal{L}\), we identify two relevant subsets:
a calibration subset \(\mathcal{L}_{\mathrm{cal}} \subset \mathcal{L}\), 
used to determine the correction factors \(\Lambda_{\mathrm{opt}}(\lambda)\),
and an extrapolation subset 
\(\mathcal{L}_{\mathrm{ext}} := \mathcal{L} \setminus \mathcal{L}_{\mathrm{cal}}\),
for which the extrapolated values of \(\Lambda_{\mathrm{opt}}(\lambda)\) are applied.
The calibration subset can be conveniently written as 
$\mathcal{L}_{\mathrm{cal}} = \{\lambda_1, \lambda_2, \dots, \lambda_{r_0}\},$
where the integer \(r_0 < r\) denotes the number of elements used for calibration. If the measurements lead to an imprecise function \(\tilde{f}(\lambda,\nu_i(\lambda))\), we can define a global correction factor \(\Lambda(\lambda)\). To do so, we introduce the deviation functional 
\begin{equation} 
\mathcal{E}(\lambda) := \sum_{\nu_i(\lambda) \in \mathcal{V}(\lambda)}
\bigl|\Lambda(\lambda)\tilde{f}(\lambda,\nu_i(\lambda)) - f(\lambda,\nu_i(\lambda))\bigr|.
\end{equation}
To correct every \(\tilde{f}(\lambda,\nu_i(\lambda))\) as much as possible, the sum above should be minimized.
In this scenario, we define \(\Lambda_{\text{opt}}(\lambda)\) as the optimized factor that delivers the minimized term \(\mathcal{E}_{\text{min}}(\lambda)\):
\begin{equation}
\label{error_min}
\mathcal{E}_{\text{min}}(\lambda) =
\sum_{\nu_i(\lambda) \in \mathcal{V}(\lambda)}
\bigl|\Lambda_{\text{opt}}(\lambda)\tilde{f}(\lambda,\nu_i(\lambda)) - f(\lambda,\nu_i(\lambda))\bigr|.
\end{equation}
It is important to notice that \(\tilde{f}(\lambda,\nu_i(\lambda))\) is hardware-dependent, which means the correction factor \(\Lambda_{\text{opt}}(\lambda)\) equally depends on the hardware configuration. Therefore, all calibrations and extrapolations must be performed on the same device under consistent conditions. Also, even though the minimized value in Eq.~(\ref{error_min}) itself is not relevant, the optimized factor \(\Lambda_{\text{opt}}(\lambda)\) is. 
We can define the refined quantity
\begin{equation}
\label{f_opt}
    \tilde{f}_{\text{opt}}(\lambda,\nu_i(\lambda)) := \Lambda_{\text{opt}}(\lambda)\tilde{f}(\lambda,\nu_i(\lambda)),
\end{equation}
which provides the mitigated version of the noisy data. 
Since \(\Lambda_{\text{opt}}(\lambda)\) depends on \(\lambda\), numerical estimates of it can be made for each 
\(\lambda \in \mathcal{L}_{\mathrm{cal}}\), and an extrapolation can then be constructed for the remaining 
\(\lambda \in \mathcal{L}_{\mathrm{ext}}\). 
This procedure allows predicting \(\Lambda_{\text{opt}}(\lambda)\) for any 
\(\lambda \in \mathcal{L}\) without requiring the analytical reference values \(f(\lambda,\nu_i(\lambda))\) to compute it explicitly.
Additionally, after we calculate \(\tilde{f}(\lambda,\nu_i(\lambda))\) from quantities obtained in the quantum circuit, this extrapolated factor \(\Lambda_{\text{opt}}(\lambda)\) can naturally be used to improve this noisy function by making use of Eq.~(\ref{f_opt}). More broadly, although presented here in the specific context of the PIED algorithm, the Correction Factor Extrapolation (CFE) method relies only on analytical relations across families of related quantum circuits and does not depend on number-theoretic structure. As such, it is applicable to a broader class of quantum algorithms that extract structured information from expectation values or dynamical observables, particularly in the NISQ regime.

In the case of our demonstrations, all these terms introduced in this technique are well known. We have:
\begin{align}
    \lambda &:= d, \\
    \nu_i(\lambda) &:= n, \\
    f(\lambda,\nu_i(\lambda)) &:= \alpha_n.
\end{align}
Then, since \(d\) must be a power of two and our interval of interest is \(\mathcal{N}_d = [2,2(d-1)]\), we define
\begin{equation}
\mathcal{L} := \{4,8,16,\dots,2^r\},
\end{equation}
and
\begin{equation}
\mathcal{V}(\lambda) := \mathcal{N}_d.     
\end{equation}
Obtaining a good extrapolation with a reasonable amount of implementations of our algorithm using this technique means we must try a few number of \(r_0\) values. We find \(r_0 = 3\) as a fair value for the extrapolation of \(\Lambda_{\text{opt}}(\lambda)\) for the next number of dimensions. That is, the extrapolation of this factor in our implementations was done using \(d = 4, 8\) and \(16\).

\subsection{Implementation of PIED with CFE}
\label{sec:implementation}
The hardware implementations we will present here were executed on IBM quantum hardware, using the library Qiskit~\cite{qiskit} (version 2.1.0) with Python (version 3.9.19). All the implementations shown in this section were performed on the Heron r2 processor Aachen, for three choices of dimension \(d\): \(4, 8\) and \(16\). For each dimension, respectively, we used \(p = 30\), \(p = 120\) and \(p = 480\), while fixing \(2^{13}\) shots and \(\omega = 0.1s^{-1}\) for all cases. For dimensions \(d = 4\) and \(d = 8\), we performed three batches of implementations with associated error bars, while for \(d = 16\) we implemented one batch, as the Aachen processor became unavailable in recent changes made to IBM's platform, which limited us to having only one batch and no error bars. To obtain the Fourier modes \(\alpha_n\), we used Simpson's rule via the Scipy library (version 1.13.1). Using the aforementioned post-processing method, numerical tests applied to the implementations led us to obtain \(\Lambda_{\text{opt}}(4) \approx 1.2678\), \(\Lambda_{\text{opt}}(8) \approx 1.7359\) and \(\Lambda_{\text{opt}}(16) \approx 1.7172\). The reduced purity curves and Fourier modes corresponding to the theory and hardware implementations are shown in Fig.~\ref{fig:purity_fourier}, respectively, in red and blue colors, along with the mitigated results in orange. 

\begin{figure*}
    \centering
    \includegraphics[width=0.9\textwidth]{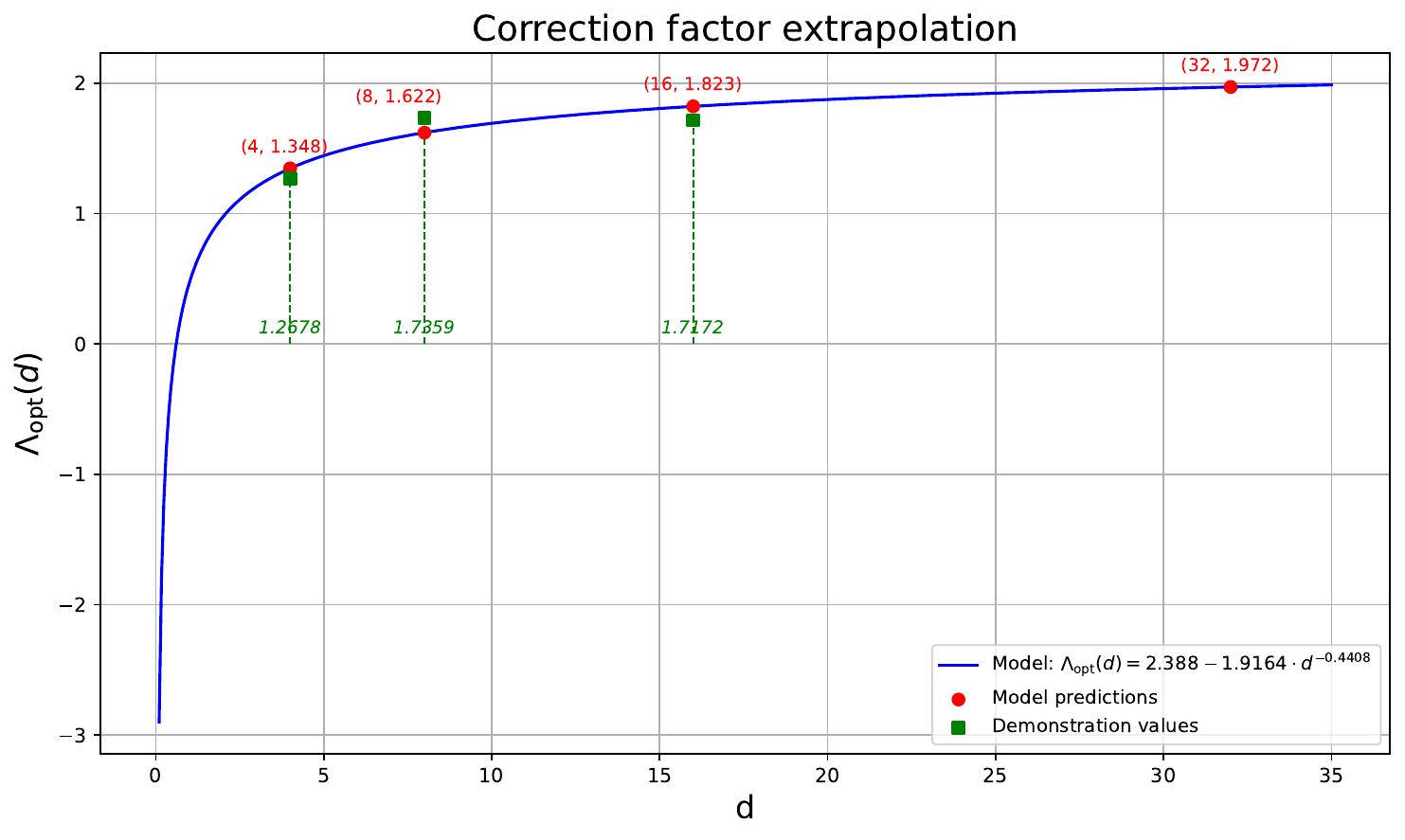}

    \vspace{1ex}
    \refstepcounter{figure}%
    \begin{minipage}[t]{0.9\textwidth}
        \justifying
        FIG.~\thefigure. Extrapolation of the correction factor \(\Lambda_{\text{opt}}(d)\) based on results from the implementations for \(\text{d} = 4, 8\) and \(16\). The green dots represent the numerically determined optimal correction factors for the Fourier mode data obtained from the hardware implementations. The blue curve corresponds to the constructed model \(\Lambda_{\text{opt}}(d)\), which predicts the correction factor behavior for any dimension \(\text{d} \geq 4\). Red dots, shown in the tuple format \(\bigl(d, \Lambda_{\text{opt}}(d)\bigr)\), are sampled directly from the model and lie on the blue curve.
    \end{minipage}
    \label{fig:correction_factor}
\end{figure*}

In order to model the correction factor \(\Lambda_{\text{opt}}(d)\), we develop some arguments to restrict the possibilities of extrapolation functions. First, this function should be strictly increasing in the dimension \(d\). Reasonably, the rate of growth of this function should be decreasing in \(d\), as we do not expect it to increase without bounds, considering the noise in the quantum computer used. This means its second derivative should be negative for any \(d \geq 4\), with its value approaching zero as we increase \(d\). Finally, introducing an asymptotic converging value, defined as \(\Lambda_0\), we can guarantee the function stabilizes. Using these arguments, 
we model the correction factor using the functional form
\begin{equation}
\Lambda_{\text{opt}}(d) = \Lambda_0 + \kappa d^{\eta}.
\end{equation}
Applying and optimizing this relation according to our results, the correction factor \(\Lambda_{\text{opt}}(d)\) respects the following expression:
\begin{equation}   
\Lambda_{\text{opt}}(d) = 2.388 -1.9164d^{-0.4408}.
\end{equation}
The curve in Fig.~\ref{fig:correction_factor} confirms that the optimized correction factor \(\Lambda_{\text{opt}}(d)\) increases monotonically with the system dimension \(d\), as expected from our analysis. Its asymptotic behavior reveals a clear tendency to saturate at \(\Lambda_0 \approx 2.388\), indicating that the rescaling strength required for mitigation gradually stabilizes as the system size grows. The value obtained for \(d = 16\) appears slightly below the fitted trend, which can be attributed to the absence of error bars in this point—stemming from the limited number of repetitions used in the corresponding implementations, as discussed previously. Overall, the data show excellent agreement with the proposed functional form, reinforcing the consistency of the extrapolation model.
\section{Conclusions}
\label{sec:conclusions}

Concluding, we have presented in this article the PIED (Prime Identification via Entanglement Dynamics) algorithm for prime number identification, previously shown to be potentially efficient~\cite{santos2024}. We implemented the PIED algorithm on IBM Quantum processors, primarily on the Aachen device, for dimensions \( d = 4, 8, \) and \( 16 \). By analyzing the time evolution of entanglement in a bipartite system, we extracted Fourier components that act as signatures of primality. Within the interval \( \mathcal{N}_d = [2, 2(d-1)] \), PIED consistently identifies prime numbers by associating them with minimal values of the corresponding Fourier modes, while composite numbers yield larger and distinguishable values.

Furthermore, we extended the theoretical analysis of these Fourier modes and established in Appendix \ref{appendix:composite_bound} that, for most composite numbers \( n \), a lower bound \( P_n \) must be respected. This bound arises from the number of divisors of \( n \), which directly influences the corresponding Fourier modes. We rigorously identified the rare exceptions to this bound, which occur for certain semiprimes involving the factors \( 2 \) and \( 3 \) within specific subintervals of \( \mathcal{N}_d \). This refined understanding allows us to define a reliable tolerance window when running PIED on quantum hardware, accounting for the discrepancies between theoretical predictions and the noisy results from implementations.

In addition to the theoretical developments, this work introduced and validated a specific noise-mitigation strategy, the Correction Factor Extrapolation (CFE) method. CFE is based on a global rescaling of the Fourier modes and requires only a single set of circuit executions for each chosen system parameter. It constructs an optimized correction factor \( \Lambda_{\text{opt}} \) that can be extrapolated across different configurations, offering a computationally inexpensive alternative to traditional error-mitigation techniques. Additional zero-noise extrapolation (ZNE) results, presented in Appendix \ref{appendix:zne}, illustrate how CFE achieves comparable or superior accuracy with substantially lower resource overhead, alongside recent learning-based approaches~\cite{czarnik2025}, underscoring its practicality for NISQ-era quantum devices. 

We also explored an alternative class of initial states, spin coherent states~\cite{arecchi1972,radcliffe1971,perelomov1986}, as detailed in Appendix \ref{appendix:spin}. Their efficient preparation becomes increasingly challenging for large \( d \), but numerical simulations for \( d = 4 \) and \( d = 8 \) indicate that the number of time partitions \( p \) required for Fourier integration can scale more favorably with \( d \) than in the case of uniform superposition states. Although these results are preliminary and limited to small dimensions, they suggest that appropriately chosen initial states might help reduce the computational effort in certain implementations without altering the algorithmic principles of PIED. 

The overall computational cost of the present implementation of PIED is dominated by the numerical extraction of Fourier components from discretely sampled values of the entanglement function. While the circuit depth associated with state preparation, time evolution, and measurement scales polynomially with the number of qubits $q = \log_{2} d$, the number of time samples required to reach a fixed numerical accuracy is determined mainly by the temporal smoothness of the entanglement function.

For the uniform superposition state employed in the main text, our numerical results indicate that achieving a given precision requires a number of sampling points scaling as $\mathcal{O}(d^{2})$. This scaling is not claimed to be optimal and should not be regarded as intrinsic to the algorithmic structure itself. As illustrated by the simulations for spin coherent states reported in Appendix \ref{appendix:spin}, alternative choices of initial states can noticeably smooth the time dependence of the entanglement function, leading to an empirical reduction of the dominant sampling cost from $\mathcal{O}(d^{2})$ to approximately $\mathcal{O}(d)$ at comparable accuracy. These results indicate that the leading source of unfavorable scaling can be circumvented within the same algorithmic framework, and they motivate further work on optimizing both state preparation and Fourier-mode extraction strategies.

Altogether, our results validate PIED’s robustness under realistic hardware conditions and highlight the broader potential of combining analytical insights with hardware-based implementations. The combination of entanglement-based observables, resource-efficient noise-mitigation strategies such as CFE, and complementary approaches to state preparation, including spin coherent states, points to a broad landscape for future work, such as optimizing circuit architectures, exploring hybrid mitigation frameworks, and extending the algorithm toward other number-theoretic or spectral applications.

\begin{acknowledgments}
This work was supported by the Coordination for the
Improvement of Higher Education Personnel (CAPES),
Grant No. 23081.002220/2025-87, by the National Council for Scientific and Technological Development (CNPq), Grants No. 300083/2025-4, 162791/2025-9, No. 132266/2025-3, No. 409673/2022-6, and No. 421792/2022-1, and the National   Institute for the Science and Technology of Applied Quantum Computing (INCT-CQA), Grant No. 408884/2024-0.
\end{acknowledgments}

\textbf{Data availability.} The data that support the findings of this study are openly available~\cite{github}.

\appendix

\section{Absolute bound for composite \(n\)}
\label{appendix:composite_bound}

\begin{figure*}[t]
    \centering
    \includegraphics[width=1\textwidth]{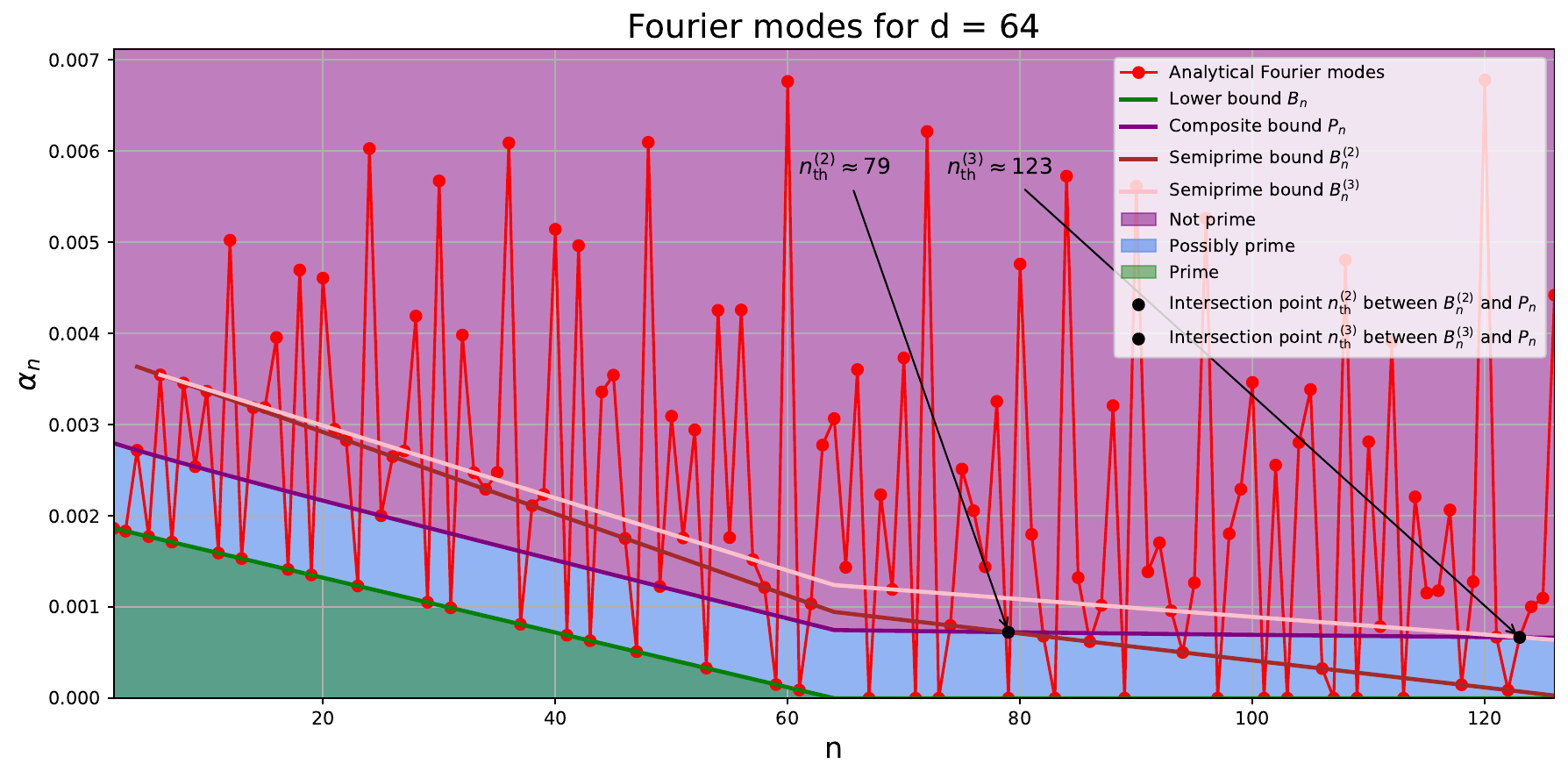}

    \vspace{1ex}
    \refstepcounter{figure}%
    \begin{minipage}[t]{0.95\textwidth}
        \justifying
        FIG.~\thefigure. Analytical Fourier modes for \(\text{d} = 64\), showing that the composite bound \(P_n\) may be violated after the threshold points \(n_{\text{th}}^{(2)} \approx 79\) and \(n_{\text{th}}^{(3)} \approx 123\). The brown and pink lines represent the interpolated Fourier modes corresponding to semiprimes with \(k = 2\) and \(k = 3\), respectively. Each of these lines eventually reaches the composite bound \(P_n\); the larger the value of \(k\), the higher the value of \(n\) at which this occurs. In this case, both \(B_n^{(2)}\) and \(B_n^{(3)}\) intersect \(P_n\) within the interval \(\mathcal{N}_d\), illustrating the coexistence of the two possible reversals of inequality~(\ref{Pnineq}). The light blue region between the composite bound \(P_n\) and the lower bound \(B_n\) indicates where a divisibility test must be applied—dividing \(n\) by \(\sqrt{n}\), and possibly also by \(2\) or \(3\), depending on the subinterval to which \(n\) belongs. Due to noise in the hardware implementations, the region above \(P_n\) is shaded purple to indicate that the corresponding values of \(n\) are confidently classified as composite. Conversely, the green region below \(B_n\) represents values of \(n\) that are certainly prime.
    \end{minipage}
    \label{fig:analytical_fourier}
\end{figure*}

Here, we obtain a new theoretical result, crucial for implementation of the PIED algorithm on quantum hardware. To motivate the development of this result, it is necessary to consider that quantum computing is not yet fault tolerant. This opens up the possibility of relevant errors in our implementations, which inevitably leads to a complication in the conclusions permitted by our algorithm. Thus, since we have the lower bound \(B_n\) of Fourier modes for prime numbers, it is almost obligatory to obtain a lower bound also for composite numbers. This allows us to define a tolerance when analyzing results gained in hardware implementations and transform the algorithm into a more reliable tool for prime number identification.

Now we focus on establishing this new bound for composite \(n\), specific to the case where the initial state is the uniform superposition, i.e., \(c_j = \frac{1}{\sqrt{d}}\) for any \(1 \leq j \leq d\). In this setting, we prove that for most composite integers \( n \), the Fourier mode \( \alpha_n \) is lower bounded not only by the trivial bound \( B_n \), but also by a tighter second bound \( P_n > B_n \), for \(n = k^2\) and the prime \(k\). We will show here that possible exceptions to this refined bound can occur only for semi-primes of \(2\) and \(3\), i.e., \(n = 2k\) or \(n = 3k\). We will denote by \(\mathcal{N}_d\) the interval of integer numbers \(n\):
\begin{equation}
    \mathcal{N}_d := [2,2(d-1)].
\end{equation}

Firstly, the contribution from the trivial divisors yields
\begin{equation}
    B_n = \frac{8}{d^4} (d - n)(d - 1),
\end{equation}
valid for \( 2 \leq n \leq d - 1 \), and set to zero for \( n \geq d \). Consequently, the expression for \( \alpha_n \) under this state becomes
\begin{align}
\alpha_n &= B_n + 4 \sum_{s=2}^{z-1} \sum_{k=1}^{d - \frac{n}{y^{(n)}_s}} \sum_{m=1}^{d - y^{(n)}_s} \frac{1}{d^4} \notag \\
&= B_n + \frac{4}{d^4} \sum_{s=2}^{z-1} \left(d - \frac{n}{y^{(n)}_s}\right)\left(d - y^{(n)}_s\right) \notag \\
&= B_n + \frac{4}{d^4} \sum_{s=2}^{z-1} \left(d^2 + n - d \biggr( y^{(n)}_s + \frac{n}{y^{(n)}_s} \biggr) \right).
\end{align}
It is convenient to define the auxiliary function
\begin{equation}
Y_n(k) := k + \frac{n}{k},
\end{equation}
motivated by the appearance of the term \( y^{(n)}_s + \frac{n}{y^{(n)}_s}\) in the expression for \(\alpha_n\). In the case \( z = 4 \), it is possible to compare \(Y_n(k)\) values for every value of \(k \leq d - 1\). For a composite \(n\) to be divisible by some \(k\) in this range, it must at least be divisible by \(2\). Therefore, we consider only \(n \geq 2k \) for this comparison. Of course, we must have 
\begin{equation}
\label{Yineq}
    Y_n(2) \geq Y_n(k),
\end{equation}
with equality iff \(n = 2k\). Now, to facilitate, we will introduce a notation for the Fourier modes \(\alpha_n\) associated with values of \(n\) with \(z = 4\) distinct divisors and a nontrivial divisor \(k\). That is, we define
\begin{equation}
\label{Bk}
    B_n^{(k)} := B_n + \frac{8}{d^4}  \left(d^2 + n - d \left( k + \frac{n}{k} \right) \right).
\end{equation}
Naturally, inequality (\ref{Yineq}) leads to
\begin{equation}
B_n^{(k)} \geq B_n^{(2)}.   
\end{equation}
From this, we conclude, at least for values of \(n \geq 2k\) with \(z \geq 4\), that
\begin{equation}
\label{B2ineq}
\alpha_n \geq B^{(2)}_n.
\end{equation}
However, the picture changes when \(z = 3\). Evidently, in these cases, \( n = k^2 \), for prime \(k\), and the only nontrivial divisor is \(k = \sqrt{n}\). The respective Fourier modes, defined as 
\begin{equation}
P_n := \alpha_{n = k^2} = B_n + \frac{4}{d^4} \left( d^2 + n - 2d \sqrt{n} \right),
\end{equation}
introduces a stronger composite \(n\) bound. For a substantial initial segment of integers \(n \in  \mathcal{N}_d\), it is true that
\begin{equation}
\label{Pnineq}
B_n^{(k)} > P_n.  
\end{equation} 
Since our goal is to obtain a global minimum of \(\alpha_n\) over composite values of \(n\), it is necessary to analyze the eventual reversal of inequality (\ref{Pnineq}). 

We begin by noting that the maximum element of the set \( \mathcal{N}_d \) is bounded above by \( 2(d - 1) \). At the same time, the product of the two smallest prime numbers greater than \( d/2 \) already exceeds this bound. Consequently, any semiprime formed solely from such primes necessarily lies outside \( \mathcal{N}_d \). That means we should only consider
\begin{equation}
k < \frac{d}{2}.
\end{equation}
In this context, we now aim to find the point where the eventual reversal of inequality (\ref{Pnineq}) may happen. For that matter, we define
\begin{align}
    \Delta(n,k) &:= \frac{d^4}{4}\bigl(B_n^{(k)} - P_n\bigr) \notag \\
    &= d^2 + n -2d\bigl(k + \frac{n}{k}\bigr) + 2d\sqrt{n},
\end{align}
and check when \(\Delta(n,k)\) begins to be negative. The expression for \(\Delta(n,k)\) has only one critical point in \( n \), which is a maximum at \( n_{\text{crit}} = \frac{d^2k^2}{(2d - k)^2}\). Therefore, the inequality (\ref{Pnineq}) is violated at possibly two threshold values \(n_{\pm}^{(k)}\), i.e., 
\begin{equation}
    \Delta(n_{\pm}^{(k)}, k) = 0,
\end{equation}
given by \(n_{\pm}^{(k)} = \frac{\frac{4d^3}{k} - 6d^2 + 4dk \pm 4\sqrt{\frac{2}{k}}(d-k)d^{3/2}}{2(1-\frac{2d}{k})^2}\). The left root, \(n_{-}^{(k)}\), is not possible for the range of values \(k < \frac{d}{2}\). To see that, we calculate \(\Delta(n,k)\) and \(\frac{\partial\Delta(n,k)}{\partial_n}\) both at \(n = 0\). We have
\begin{equation}
    \Delta(0,k) = d^2 -2dk > 0,
\end{equation}
and
\begin{equation}
    \frac{\partial\Delta(n,k)}{\partial_n} = 1 -\frac{2d}{k} + \frac{d}{n} \implies \frac{\partial\Delta(0,k)}{\partial_n} = +\infty.
\end{equation}
Considering \(\Delta(n,k)\) is only defined for \(n \geq 0\), these calculations allow us to conclude that the only admissible threshold value \(n_{\text{th}}^{(k)}\) is the rightmost root, \(n_{+}^{(k)}\):
\begin{equation}
\label{n_th}
n_{\text{th}}^{(k)} =
\frac{
\frac{4d^3}{k}
- 6d^2
+ 4dk
+ 4\sqrt{\frac{2}{k}}\,(d-k)\,d^{3/2}
}{
2\left(1-\frac{2d}{k}\right)^2
}.
\end{equation}
Although this expression has a complicated appearance, it is possible to extract analytical information about how the threshold value \(n_{\text{th}}^{(k)}\) behaves as \(k\) varies. In particular, we are interested in verifying that \(n_{\text{th}}^{(k)}\) increases monotonically with \(k\), since this guarantees that the point where inequality~(\ref{Pnineq}) reverses occurs later for larger \(k\). To verify that the derivative of Eq.~(\ref{n_th}) with respect to \( k \) is positive for all \( k < d/2 \), we set \( k = x d \), with
\begin{equation}
x < \frac{1}{2}.
\end{equation}
Substituting this into Eq.~(\ref{n_th}) and simplifying gives
\begin{equation}
n_{\text{th}}^{(x)} = d^2 F(x),
\end{equation}
where
\begin{equation}
F(x)
=
\frac{
-2\sqrt{2}\,x^{5/2}
+ 2\sqrt{2}\,x^{3/2}
+ 2x^3
- 3x^2
+ 2x
}{
(x-2)^{2}
}.
\end{equation}
Differentiating \( F(x) \) and simplifying, we obtain
\begin{equation}
\begin{split}
\frac{\partial F}{\partial x}
&=
\frac{
- \sqrt{2}\,x^{5/2}
+ 9\sqrt{2}\,x^{3/2}
- 6\sqrt{2}\,x^{1/2}
+ 2x^3
}{\left(x - 2\right)^3}
\\
&\quad
+
\frac{
- 12x^2
+ 10x
- 4
}{
\left(x - 2\right)^3
}.
\end{split}
\end{equation}
The numerator and denominator are both negative, so \( \frac{\partial F}{\partial x} > 0 \) in this range. Since \( n_{\text{th}}^{(x)} = d^2 F(x) \) with \( d^2 > 0 \), we also have
\begin{equation}
\frac{\partial n_{\text{th}}^{(x)}}{\partial x} > 0
\quad \text{for all} \quad
x < \frac{1}{2}.
\end{equation}
Finally, because \( x = k/d \), monotonicity in \( x \) is equivalent to monotonicity in \( k \). Thus,
\begin{equation}
\frac{\partial n_{\text{th}}^{(k)}}{\partial k} > 0
\quad \text{for all} \quad
k < \frac{d}{2}.
\end{equation}
Consequently, \(n_{\text{th}}^{(k)}\) is a strictly increasing function of \(k\) in this region. This implies that for any positive integer \(h\),
\begin{equation}
n_{\text{th}}^{(k)} < n_{\text{th}}^{(k+h)},
\end{equation}
meaning that each \(B_n^{(k)}\) reverts inequality~(\ref{Pnineq}) at a smaller value of \(n\) than \(B_n^{(k+h)}\) does. Naturally, there will be a point where \(n_{\text{th}}^{(k)}\) is outside the interval \(\mathcal{N}_d\). Thus, there is a maximum value integer value \(k_{\text{max}} < \frac{d}{2}\) where the inequality reversal is possible. This permits us to infer that, for a given \(d\), the violation of inequality (\ref{Pnineq}) can only occur for the set of prime numbers
\begin{equation} 
\label{Kset}
\mathcal{K}_{d}:= \{2,3,\dots,k_{\text{max}}\}.
\end{equation}
In fact, it turns out that this set only contains at most \(k = 2\) and \(k = 3\), as we will prove now.

Considering that for some \(k\) the inequality can be violated for a subset of the interval \( \mathcal{N}_d\), the function \(B_n^{(k)}\) crosses the curve of \(P_n\) at some point \(n_{\text{th}}^{(k)}\), which implies \(P_{n_{\text{th}}} = B_{n_{\text{th}}}^{(k)}\). Then, any point \(n_0\) such that \(2(d-1) \geq n_0 \geq n_{\text{th}}^{(k)}\) should correspond to \(P_{n_0} \geq B_{n_0}^{(k)}\). Therefore, it is guaranteed that the point \(2(d-1)\) should satisfy
\begin{equation}
P_{2(d-1)} \geq B_{2(d-1)}^{(k)}.    
\end{equation}
Mathematically, this inequality can be expressed as \(0 \geq \Delta(2(d-1),k)\). Developing it further, we have:
\begin{align}
\label{kineq}
& 0 \geq \Delta(2(d-1),k), \\
& 0 \geq d^2 + 2(d-1) - 2d\left(k + \frac{2(d-1)}{k}\right) + 2d\sqrt{2(d-1)},  \notag \\
& 0 \geq -2dk^2 +  \left(d^2 + 2(d-1) + 2d\sqrt{2(d-1)} \right)k -4d(d-1). \notag
\end{align}
Then, we obtain two roots:
\begin{align}
\label{k-}
k_{-} &= \frac{\bigl(d + \sqrt{2(d-1)}\bigr)^2 - u}{4d}, \\
k_{+} &= \frac{\bigl(d + \sqrt{2(d-1)}\bigr)^2 + u}{4d},  
\end{align}
where \(u = \sqrt{\bigl(d + \sqrt{2(d-1)}\bigr)^4 - 32d^2(d-1)}\). This leads to the following inequalities for \(k\):
\begin{equation}
    k \leq k_{-},
\end{equation}
and
\begin{equation}
    k \geq k_{+}.
\end{equation}
Since we know that the set \(\mathcal{K}_{d}\), where inequality (\ref{kineq}) is respected, is the one defined in Eq.~(\ref{Kset}), we must choose the root \(k_{-}\) as the upper value \(k\). Evidently, \(k_{-}\) grows with \(d\), which implies that it is possible to estimate the asymptotic limit \(k_{\infty}\), and, consequently, \(\mathcal{K}_{\infty}\). To do that, we should first expand the terms inside the square root \(u\):
\begin{align}
u^2 &= \bigl(d + \sqrt{2(d-1)}\bigr)^4 - 32d^2(d-1) \notag \\
&= d^4 + 4d^3 \sqrt{2(d - 1)} - 20d^3 \notag \\
&\quad + 8d^2 \sqrt{2(d - 1)} + 24d^2 - 8d \sqrt{2(d - 1)} - 8d + 4 \notag \\
&= d^4 + 4d^3\sqrt{2(d-1)} -20d^3 + \mathcal{O}(d^{\frac{5}{2}}) \notag \\
&= d^4\biggr(1 +  \frac{4\sqrt{2(d-1)}}{d} - \frac{20}{d} + \mathcal{O}\bigl(1/d^{\frac{3}{2}}\bigr)\biggr).
\end{align}
Now, we use
\begin{equation}
    \sqrt{1+\epsilon} = 1 + \frac{1}{2}\epsilon -\frac{1}{8}\epsilon^2 + \cdots,
\end{equation}
with \(\epsilon = \frac{4\sqrt{2(d-1)}}{d} - \frac{20}{d} + \mathcal{O}\bigl(1/d^{\frac{3}{2}}\bigr)\), to write the asymptotic expression for the square root term \(u\):
\begin{align}
    u &=\sqrt{d^4\biggr(1 +  \frac{4\sqrt{2\bigl(d-1\bigr)}}{d} - \frac{20}{d} + \mathcal{O}\bigl(1/d^{\frac{3}{2}}\bigr)\biggr)}\notag \\
    &\approx \biggr[d^2 +  \frac{d^2}{2}\biggr(\frac{4\sqrt{2(d-1)}}{d} - \frac{20}{d} + \mathcal{O}\bigl(1/d^{\frac{3}{2}}\bigr)\biggr) \notag \\
    &\quad - \frac{d^2}{8}\biggr(\frac{4\sqrt{2(d-1)}}{d} - \frac{20}{d} + \mathcal{O}\bigl(1/d^{\frac{3}{2}}\bigr)\biggr)^2\biggr] \notag \\
    &= \biggr[d^2  + \biggr(2d\sqrt{2(d-1)} - 10d + \mathcal{O}\bigl(d^{\frac{1}{2}}\bigr)\biggr) \notag \\
    &\quad - \biggr(4(d-1) - 20\sqrt{2(d-1)} + 50 + \mathcal{O}(1)\biggr)\biggr] \notag \\
    &\approx \biggr(d^2 + 2d\sqrt{2(d-1)} - 14d\biggr),
\end{align}
where we have neglected terms of order \(\mathcal{O}(d^{\frac{1}{2}})\) or lower, as \(k_{-}\) contains a denominator proportional to \(d\). Finally, the asymptotic limit \(k_{\infty}\) is just
\begin{align}
    k_{\infty} &= \lim_{d \to \infty}k_{-} \notag \\
    &= \lim_{d \to \infty} \frac{\bigl(d + \sqrt{2(d-1)}\bigr)^2 - u}{4d} \notag \\
    &= \lim_{d \to \infty}\biggr(\frac{d^2 + 2d\sqrt{2(d-1)} + 2(d-1)}{4d} \notag \\
    &\quad -\frac{d^2 +2d\sqrt{2(d-1)} - 14d}{4d}\biggr)\notag \\
    &= 4.
\end{align}
This is the main theoretical result for a bound to composite integers \(n\), using the uniform superposition state. We have shown that \(\mathcal{K}_d\) may contain only prime numbers \(k = 2\) and \(k = 3\). In the asymptotic limit \(d \to \infty\), it could, in principle, also, contain \(k = 4\). However, since this particular value only satisfies
\begin{equation}
   \lim_{d \to \infty}\Delta(2(d-1),4) = 0,
\end{equation}
meaning its threshold is \(n_{\text{th}}^{(4)} = 2(d-1)\), it should not be considered. In fact, it is divisible by \(k = 2\) and is therefore not a prime number, implying that it does not belong to the set \(\mathcal{K}_{\infty}\). 

Thus, a range where \(P_n\) is the absolute minimum for composite integers can be defined. After this range, using the threshold expression (\ref{n_th}) for \(k = 2\) and \(k = 3\), it is possible to know where to start looking for composite integers that may fall below \(P_n\), that is, we know exactly when \(P_n\) loses its property of being the absolute composite \(n\) minimum. Not only that, we know that after this range any \(n\) below \(P_n\) must definitely be divisible either by \(2, 3\) or \(\sqrt{n}\). This last fact allows us to separate the interval \(\mathcal{N}_d\) into three convenient subintervals consisted of integer numbers. Formally, we define \( \mathcal{N}_d \) as the union of three disjoint open subintervals, 
together with the endpoint \( 2(d - 1) \):
\begin{align}
    \mathcal{N}_d^{(1)} &:= \bigl[2, n_{\text{th}}^{(2)}\bigr), \\
    \mathcal{N}_d^{(2)} &:= \bigl[n_{\text{th}}^{(2)}, n_{\text{th}}^{(3)}\bigr), \\
    \mathcal{N}_d^{(3)} &:= \bigl[n_{\text{th}}^{(3)}, 2(d - 1)\bigr), \\[4pt]
    \mathcal{N}_d &:= 
    \mathcal{N}_d^{(1)} \cup 
    \mathcal{N}_d^{(2)} \cup 
    \mathcal{N}_d^{(3)} \cup 
    \{2(d - 1)\}.
\end{align}
Operationally, each subinterval admits a different strategy for identifying non-prime integers \(n\) whose Fourier modes satisfies \(\alpha_n \leq P_n\):

\begin{itemize}
    \item For \(n \in \mathcal{N}_d^{(1)}\), it is sufficient to test whether \(n\) is divisible by \(\sqrt{n}\).
    \item For \(n \in \mathcal{N}_d^{(2)}\), we test for divisibility by \(\sqrt{n}\) and by 2.
    \item For \(n \in \mathcal{N}_d^{(3)}\), the test includes \(\sqrt{n}\), 2, and 3.
\end{itemize}

These theoretical results are represented as an example in Fig.~\ref{fig:analytical_fourier} for \(d = 64\). From Eq.~(\ref{n_th}), we obtain the threshold points \(n_{\text{th}}^{(2)} \approx 79\) and \(n_{\text{th}}^{(3)} \approx 123\), both lying inside \(\mathcal{N}_{64} = [2,126]\). This confirms that the intersections of \(B_n^{(2)}\) and \(B_n^{(3)}\) with \(P_n\) occur within the same interval, illustrating the two possible reversals of inequality~(\ref{Pnineq}). According to the previous discussion, values of \(n\) within the first subinterval of \(\mathcal{N}_{64}\) must be tested for divisibility by \(\sqrt{n}\) before being classified as prime. Those in the second subinterval require additional division by \(2\), and finally, elements of the third subinterval \(\mathcal{N}_{64}^{(3)}\) must be tested for divisibility by \(2\), \(3\), and \(\sqrt{n}\).

\section{PIED implementations with zero-noise extrapolation}
\label{appendix:zne}
This Appendix is organized into two sections. In the first section, we provide a brief overview of the zero-noise extrapolation (ZNE) technique, following mainly the presentation in Ref.~\cite{mitiq}. In the second section, we describe its implementation within the PIED algorithm. It is worth noting that, although our theoretical discussion of ZNE draws from the Mitiq framework, the implementation employed here was developed independently, without the use of the Mitiq library.

ZNE is an error-mitigation method used to estimate the expectation value of a quantum observable in the noiseless limit through extrapolation from measurements obtained under different noise levels. The central idea is to artificially amplify the noise in a controlled way and then extrapolate the measured observable back to the zero-noise limit. ~\cite{cai2023} This procedure enables one to infer noise-free quantities without requiring detailed modeling of the hardware noise~\cite{endo2018}.

More rigorously, given a prepared state \( \hat{\rho}(\tau) \), where \( \tau \) denotes the noise level in the quantum system, we write
\begin{equation}
    \tau_j = \xi_j \tau_0,
\end{equation}
where \( \tau_0 \) represents the intrinsic noise level of the quantum processor and \( \xi_j \ge 1 \) are noise-scaling factors. The corresponding expectation values are
\begin{equation}
    \langle \hat{O} \rangle_{\tau_j} = \mathrm{Tr}[\hat{\rho}(\tau_j) \hat{O}] = \mathrm{Tr}[\hat{\rho}(\xi_j \tau_0) \hat{O}].
\end{equation}
By obtaining \( \langle \hat{O} \rangle_{\tau_j} \) for several values of \( \xi_j \), one can fit an extrapolation curve and evaluate \( \langle \hat{O} \rangle \) at \( \tau < \tau_0 \), including the noiseless limit \( \tau = 0 \), which is the objective of the method.

Practically, the implementation of ZNE can be separated into two main steps:

\textbf{Step 1) Scaling noise:} defining a procedure to scale noise for many different values of \( \xi \) to measure \( \langle \hat{O} \rangle_\tau \);

\textbf{Step 2) Extrapolation:} fitting a curve with the data obtained in Step 1 to estimate \( \langle \hat{O} \rangle_{\tau = 0} \).

\begin{figure*}[t]
    \centering
    \makebox[\textwidth][c]{%
        \begin{subfigure}[t]{0.4\textwidth}
            \centering
            \includegraphics[width=\linewidth]{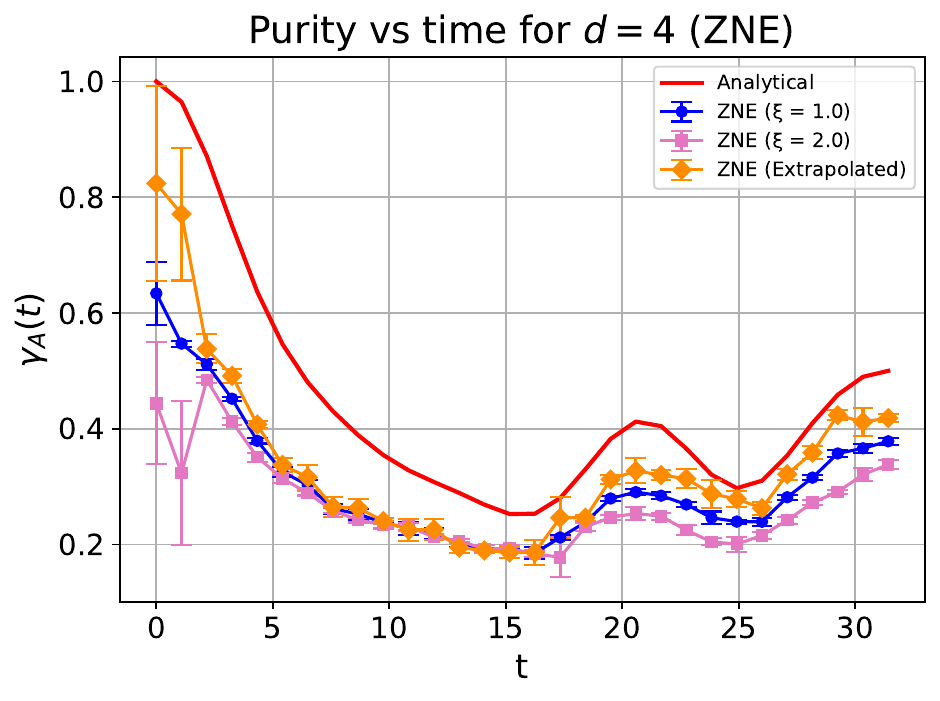}
            \caption{Time evolution of the purity of subsystem \(A\) obtained from the PIED circuit for \(d = 4\). Orange points represents the noisy data mitigated through zero-noise extrapolation (ZNE), while blue and and pink points correspond to the results at the two noise-scaling factors used in the extrapolation.}
            \label{fig:purity_zne_d4}
        \end{subfigure}
        \hspace{0.05\textwidth}
        \begin{subfigure}[t]{0.4\textwidth}
            \centering
            \includegraphics[width=\linewidth]{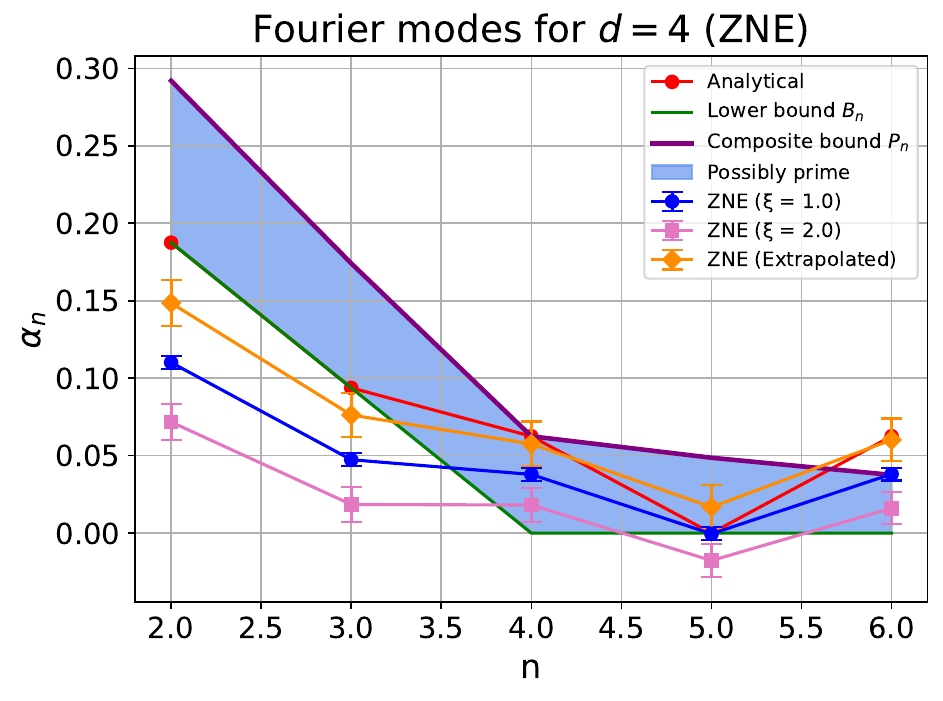}
            \caption{Fourier components extracted from the ZNE-corrected purity curve of panel~(a). The mitigated amplitudes reproduce the expected qualitative structure of the Fourier spectrum for \(d = 4\), with clear separation between the modes associated with prime and composite indices.}
            \label{fig:fourier_zne_d4}
        \end{subfigure}
    }

    \vspace{1ex}
    \addtocounter{figure}{-1}
    \refstepcounter{figure}%
    \begin{minipage}[t]{0.9\textwidth}
        \justifying
        FIG.~\thefigure. Zero-noise extrapolation (ZNE) results for the PIED algorithm with \( d = 4 \). Panel~(a) shows the evolution of the purity of subsystem \(A\) as a function of time, obtained at two scaled noise levels and extrapolated to the zero-noise limit through linear fitting. Panel~(b) displays the corresponding Fourier-mode amplitudes extracted from the mitigated data. For each point, the error bars represent the standard error of the mean over three independent batches. Specifically, for each noise-scaling factor (\( \xi = 1.0 \) and \( \xi = 2.0 \)), standard errors are computed from the three batch results, and the ZNE error bars are calculated by propagating the corresponding uncertainties from the same batch combinations used in the extrapolation. The overall agreement with the theoretical pattern demonstrates that ZNE effectively suppresses hardware noise in the estimation of observables relevant to entanglement dynamics.
    \end{minipage}

    \label{fig:purity_fourier_zne}
\end{figure*}

Enhancing noise in a quantum system can, in principle, be achieved by re-calibrating the control pulses of quantum processors, for example, reducing their amplitudes and extending their duration~\cite{znefirst}. However, such low-level control is typically inaccessible in most current quantum-computing platforms. A more practical approach, applied in this Appendix, is to scale noise digitally by increasing the circuit depth, which requires only gate-level access.

Digital noise scaling can be implemented through several methods, including unitary and layer folding (local folding), global folding, identity scaling, and parameter-noise scaling~\cite{var, he2020}. In unitary folding, each selected gate is replaced according to
\begin{equation}
    \hat{U} \rightarrow \hat{U} \hat{U}^\dagger \hat{U}.
\end{equation}
In our implementation, the noise-scaling factors are given approximately by
\begin{equation}
    \xi_j \approx 1 + \frac{2 N'_j}{N},
\end{equation}
where \( N \) denotes the total number of gates in the circuit and \( N'_j \) the number of gates subject to folding to reach the scaling factor \( \xi_j \). Consequently, \( \xi = 2 \) corresponds to folding roughly half of the gates (\( N' = N / 2 \)), while \( \xi = 3 \) corresponds to folding all gates once (\( N' = N \)). For each \( \xi_j \), every other gate of the circuit is folded in a left-to-right order until \( N' \) is reached.

Once the noise-scaling step is completed, an extrapolation curve is fitted to the measured expectation values. This can be done using several models, such as linear, polynomial, Richardson, or exponential extrapolation~\cite{digital_zne, cai2021, czarnik2021, qin2023}. The choice of model depends on the behavior of the data \( \langle \hat{O} \rangle_{\tau_j} \). In the implementations presented in Fig.~\ref{fig:purity_fourier_zne}, we adopt linear extrapolation, which estimates the zero-noise value \( \langle \hat{O} \rangle_{\tau = 0} \) through a simple linear fit.

The comparison shows that both ZNE and CFE are effective for the purpose considered here: mitigating the systematic bias in the Fourier-mode amplitudes extracted from the experimental purity data. ZNE 
can be particularly effective for individual circuit instances.
Its drawback is the additional experimental overhead required to measure the same observable at several effective noise levels. CFE is more specialized, but better aligned with the structure of PIED: it uses the family of circuits indexed by \(d\) and the available analytical reference values to transfer calibration information across dimensions. Thus, while ZNE may be preferable for isolated instances in the present setting, CFE offers a lower-overhead route that may become more advantageous for future PIED-like implementations and other structured protocols involving many related circuits.

\section{Spin coherent initial states simulations for PIED}
\label{appendix:spin}
In this Appendix, we introduce and simulate spin coherent states, focusing on using it on PIED.

\textit{Quantum state definition} -- A spin coherent state can be defined as
\begin{equation}
|\phi, \theta\rangle = e^{-i\phi \hat{J}_z}e^{-i\theta \hat{J}_y}|s,s\rangle,
\end{equation}
where $\hat{J}_z$ and $\hat{J}_y$ are angular momentum operators and $|s,s\rangle$ is the eigenvector of $\hat{J}_z$ with maximal eigenvalue $s$. It can be shown that an analytic form for this state is 
\begin{equation}
|\phi,\theta\rangle = \sum_{m = -s}^s\sqrt{
  \bigl(\begin{smallmatrix}
    2s \\
    s+m 
  \end{smallmatrix}\bigr)
}\cos(\theta/2)^{s+m}\sin(\theta/2)^{s-m}e^{-im \phi}|s,m\rangle, 
\label{general-spin-coh-state}
\end{equation}
where the notation \(\bigl(\begin{smallmatrix}
    2s \\
    s+m 
  \end{smallmatrix}\bigr)\) means 
\begin{equation}
 \bigl(\begin{smallmatrix}
    2s \\
    s+m 
  \end{smallmatrix}\bigr) = \frac{(2s)!}{(s+m)!(s - m)!}.    
\end{equation}
To prepare a general spin coherent state one can exploit the relation between the beam-splitter and angular momentum algebras~\cite{js1, js2}. To show this correspondence, we begin with the unitary
\begin{equation}
\hat{U}_{BS}(\phi, \theta)=\exp\left[\frac{\theta}{2}\left(\hat{a}_0^\dagger \hat{a}_1e^{i\phi}-\hat{a}_0\hat{a}_1^\dagger e^{-i\phi}\right)\right],
\label{BS}
\end{equation}
where $\theta/2=\arctan(r/t)$ relates the reflection $r$ and transmission $t$ coefficients, $\phi$ is the phase difference between the reflected and transmitted beams and $\hat{a}_0, \hat{a}_1$ ($\hat{a}_0^\dagger, \hat{a}_1^\dagger$) are the boson annihilation (creation) operators of input modes 0 and 1, respectively. It can be shown by the Baker-Hausdorff lemma that the input and output mode operators relate through 
\begin{equation}
\begin{aligned}
\hat{a}_0&=\cos(\theta/2)\hat{a}_2-e^{i\phi}\sin(\theta/2)\hat{a}_3 \\  
\hat{a}_1&=\cos(\theta/2)\hat{a}_3+e^{-i\phi}\sin(\theta/2)\hat{a}_2,
\end{aligned}
\end{equation}
where $\hat{a}_2, \hat{a}_3$ ($\hat{a}_2^\dagger, \hat{a}_3^\dagger$) are the annihilation (creation) operators of output modes 2 and 3. By acting (\ref{BS}) on the state $|0_0, N_1\rangle$ (vacuum on mode $0$, $N$ bosons on mode $1$) one gets (in the Heisenberg picture):
\begin{equation}
\begin{aligned}
&|0_0, N_1\rangle =\frac{(\hat{a}_1^\dagger)^N}{\sqrt{N!}}|0_0, 0_1\rangle \\ &=\frac{1}{\sqrt{N!}}\left(e^{i\phi}\sin(\theta/2)\hat{a}_2^\dagger +\cos(\theta/2)\hat{a}_3^\dagger \right)^N|0_2, 0_3\rangle \\ &=\frac{1}{\sqrt{N!}}\sum_{k=0}^N\binom{N}{k}\big(e^{i\phi}\sin(\theta/2)\hat{a}_2^\dagger\big)^k\big(\cos(\theta/2)\hat{a}_3^\dagger\big)^{N-k}|0_2, 0_3\rangle \\ &=\frac{1}{\sqrt{N!}}\sum_{k=0}^N\binom{N}{k}\big(e^{i\phi}\sin(\theta/2)\big)^k\cos(\theta/2)^{N-k}\\  &\hspace{60pt}\times\sqrt{k!(N-k)!}|k_2, (N-k)_3\rangle \\ &=\sum_{k=0}^N \sqrt{\binom{N}{k}}\sin(\theta/2)^k\cos(\theta/2)^{N-k} \\&\hspace{60pt}\times e^{ik\phi}|k_2, (N-k)_3\rangle.
\end{aligned}
\end{equation}

Applying the Jordan-Schwinger state map 
\begin{equation}
 |k, N-k \rangle \mapsto |s+m, s-m\rangle,   
\end{equation}
we find that the action of $\hat{U}_{BS}$ on the state $|0, N\rangle$ generates a state that can be mapped into $|\phi, \theta\rangle$, exactly as it is defined in (\ref{general-spin-coh-state}). To do that, we set $\theta \to \pi - \theta$ and $\phi \to -\phi$~\cite{schwinger}. Concisely, 
\begin{equation}
\hat{U}_{BS}(-\phi, \pi-\theta)|0, N=2s \rangle \mapsto |\phi, \theta \rangle.
\end{equation}
The implementation of the beam-splitter unitary is well known in the literature~\cite{bs1, bs2, bs3}. The process involves mapping bosonic operators to qubit operators and decomposing the unitary into quantum logic gates. Although simple in concept, the
complexity of this implementation grows rapidly with increasing $N$, caused by the increase of both number of gates and number of decomposition steps (e.g. Trotter steps), making it impractical in current hardware. With that in mind, we have used an heuristic approach for preparing such states in this Appendix. 

\textit{State preparation} -- We prepare a product state \(|\psi(0)\rangle_{AB}\) for a bipartite system \(AB\), where each subsystem (\(A\) and \(B\)) is initialized in an identical spin coherent state with \(\theta = \frac{\pi}{2}\) and \(\phi = 0\). Using these values, the individual state reduces to
\begin{equation}
\label{subsystem-spin-coh-state}
    |0,\frac{\pi}{2}\rangle_{S} = \left(\frac{1}{2^s}\right)\sum_{m = -s}^s\sqrt{
  \bigl(\begin{smallmatrix}
    2s \\
    s+m 
  \end{smallmatrix}\bigr)
}|s,m\rangle,
\end{equation}
where \(S \in \{A,B\}\). Therefore, the total state is given by the tensor product of these two spin coherent states:
\begin{equation}
\label{AB-spin-coh-state}
    |\psi(0)\rangle_{AB} =\left(\frac{1}{4^s}\right)\!\sum_{m_{A}, m_{B} = -s}^s\sqrt{
  \bigl(\begin{smallmatrix}
    2s \\
    s+m_A 
  \end{smallmatrix}\bigr) \bigl(\begin{smallmatrix}
    2s \\
    s+m_B 
  \end{smallmatrix}\bigr)
}|m_A m_B\rangle. 
\end{equation}
In our case, \(s \in \{\frac{1}{2}, \frac{3}{2}, \frac{7}{2}, \frac{15}{2}, ...\}\), since we must have 
\begin{equation}
    d = 2s + 1,
\end{equation}
as a power of \(2\) for our algorithm. Rewriting \(m_S := \frac{d + 1}{2} - n_S\), where \(n_S \in \{1,2,\dots,d\}\), while defining the map to the computational basis \(\{|E_{n_S}\rangle\}\) as \(|m_S\rangle \mapsto |E_{n_S}\rangle\), we have:
\begin{equation}
\label{bip_spin_state}
|\psi(0))\rangle_{AB} = \frac{1}{2^{d-1}}\!\!\!\sum_{n_A, n_B = 1}^{d}\sqrt{
  \bigl(\begin{smallmatrix}
    d - 1 \\
    d - n_A 
  \end{smallmatrix}\bigr)
  \bigl(\begin{smallmatrix}
    d - 1 \\
    d - n_B 
  \end{smallmatrix}\bigr)
}|E_{n_A} E_{n_B}\rangle.
\end{equation}

\textit{PIED simulations using spin coherent states} -- The efficient preparation of state (\ref{bip_spin_state}) for arbitrary \(d\) is challenging. Thus, here we focus on doing it for \(d = 4\) and \(d = 8\). For both dimensions, we construct the quantum circuit associated with the respective quantum state preparation using Y rotations, CNOT and Hadamard gates. The preparation circuits are presented in Fig.~\ref{fig:qc_spin}. 
The circuit for \(d = 4\), shown in Fig.~\ref{fig:qc_spin_d4}, is compact, 
whereas the one for \(d = 8\) in Fig.~\ref{fig:qc_spin_d8} 
already requires additional parametrized rotations and more entangling gates.

\begin{figure*}[t]
    \centering

    \makebox[\textwidth][c]{%
        \begin{subfigure}[t]{0.3\textwidth}
            \centering
            \includegraphics[width=\linewidth]{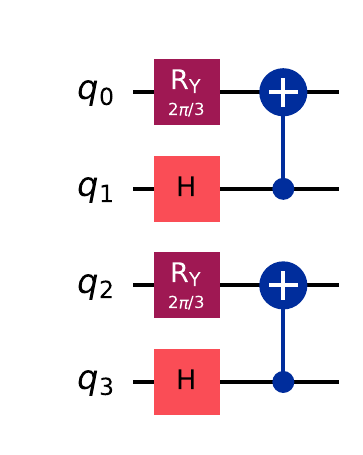}
            \caption{Circuit for \(d = 4\), using two qubits per subsystem and a combination of \(Y-\)rotations and Hadamard gates to generate the required superposition amplitudes.}
            \label{fig:qc_spin_d4}
        \end{subfigure}
        \hspace{0.05\textwidth}
        \begin{subfigure}[t]{0.45\textwidth}
            \centering
            \includegraphics[width=\linewidth]{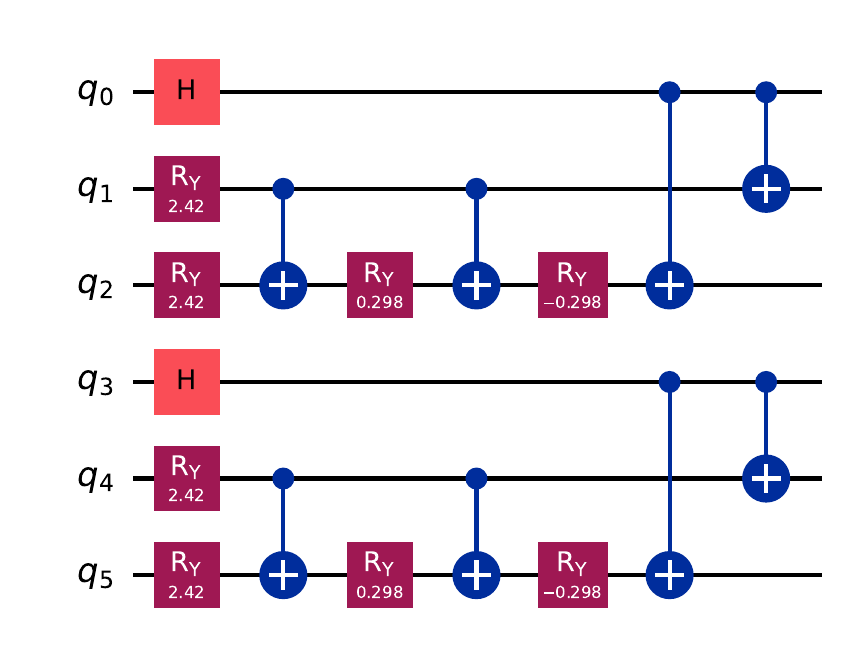}
            \caption{Circuit for \(d = 8\), which demands additional qubits and a more elaborate arrangement of parametrized \(Y-\)rotation gates to encode the target spin coherent state.}
            \label{fig:qc_spin_d8}
        \end{subfigure}
    }
    \vspace{1ex}
    \addtocounter{figure}{-1}
    \refstepcounter{figure}%
    \begin{minipage}[t]{0.9\textwidth}
        \justifying
        FIG.~\thefigure. Quantum circuits for preparing the spin coherent states as initial states in PIED implementations. These circuits implement the state given in Eq.~(\ref{AB-spin-coh-state}), corresponding to \(|0, \frac{\pi}{2}\rangle \) for both subsystems \(A\) and \(B\). The figure shows two cases, for \(d = 4\) and \(d = 8\). While the circuit for \(d = 4\) is compact and relatively simple, it does not follow a direct pattern that can be scaled up to \(d = 8\) or higher dimensions. As a result, the preparation of spin coherent states for larger \(d\) requires increasingly complex sequences of parametrized rotations and entangling operations, making scalability a significant challenge for current hardware.
    \end{minipage}

    \label{fig:qc_spin}
\end{figure*}

\begin{figure*}[t]
    \centering
    \makebox[\textwidth][c]{%
        \begin{subfigure}[t]{0.4\textwidth}
            \centering
            \includegraphics[width=\linewidth]{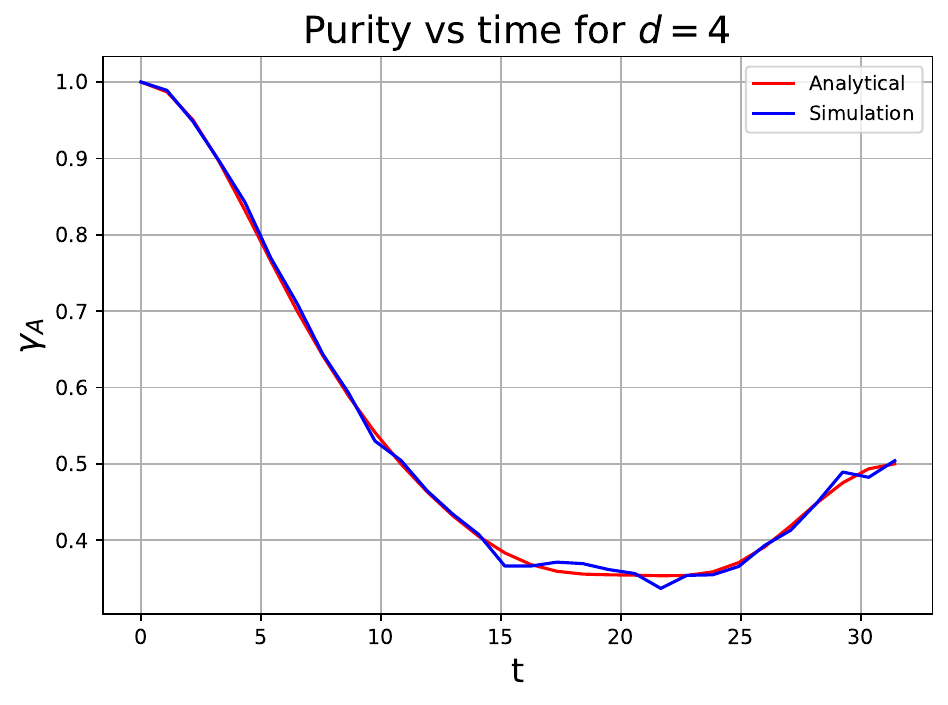}
            \caption{Time evolution of the purity of subsystem \(A\) for \( d = 4 \), obtained from numerical simulations of the PIED circuit initialized with spin coherent states.}
            \label{fig:purity_spin_d4}
        \end{subfigure}
        \hspace{0.05\textwidth}
        \begin{subfigure}[t]{0.4\textwidth}
            \centering
            \includegraphics[width=\linewidth]{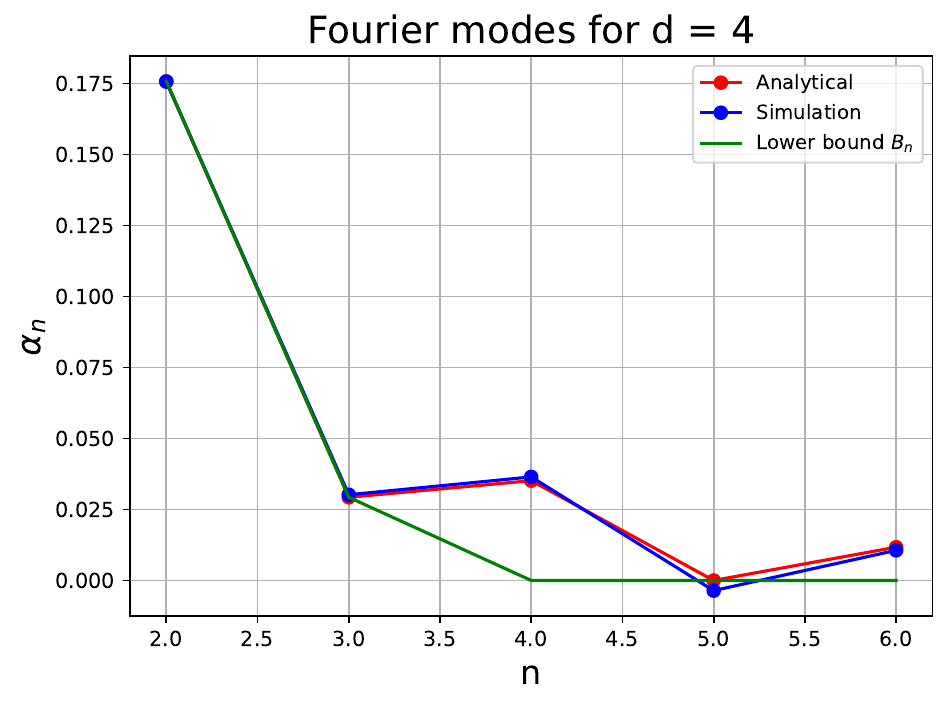}
            \caption{Fourier components extracted from the simulated purity of panel~(a) for \( d = 4 \).}
            \label{fig:fourier_spin_d4}
        \end{subfigure}
    }

    \vspace{1.2ex}

    \makebox[\textwidth][c]{%
        \begin{subfigure}[t]{0.4\textwidth}
            \centering
            \includegraphics[width=\linewidth]{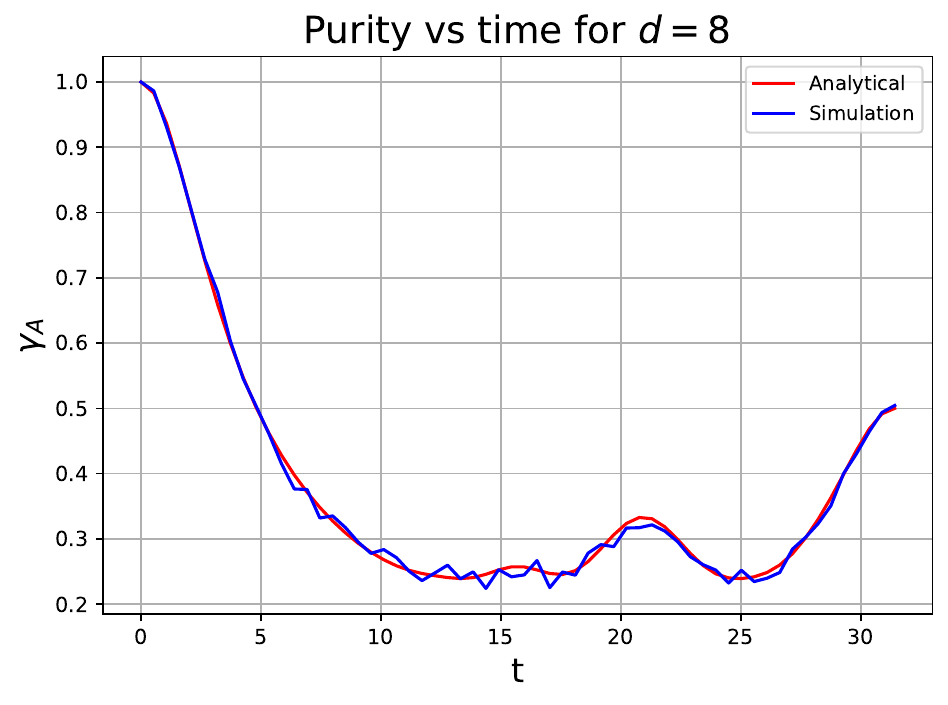}
            \caption{Time evolution of the purity of subsystem \(A\) for \( d = 8 \), obtained from numerical simulations of the PIED circuit initialized with spin coherent states.}
            \label{fig:purity_spin_d8}
        \end{subfigure}
        \hspace{0.05\textwidth}
        \begin{subfigure}[t]{0.4\textwidth}
            \centering
            \includegraphics[width=\linewidth]{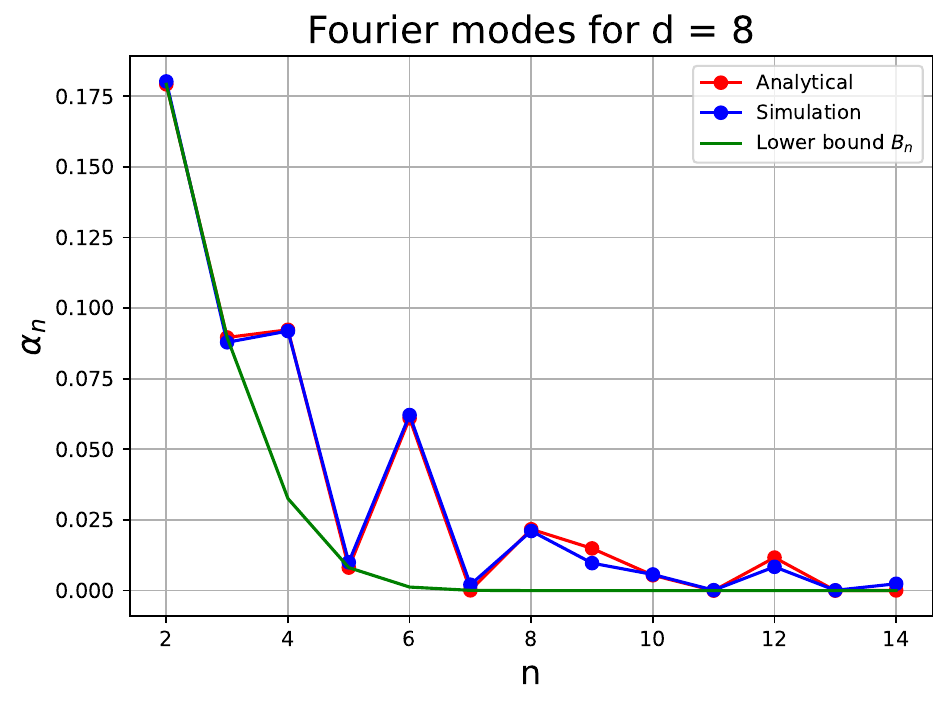}
            \caption{Fourier components extracted from the simulated purity of panel~(c) for \( d = 8 \).}
            \label{fig:fourier_spin_d8}
        \end{subfigure}
    }

    \vspace{1ex}
    \addtocounter{figure}{-1}
    \refstepcounter{figure}%
    \begin{minipage}[t]{0.9\textwidth}
        \justifying
        FIG.~\thefigure. Simulations of PIED initialized with spin coherent states for \( d = 4 \) and \( d = 8 \), performed using \(2^{13}\) measurement shots, with \(p = 30\) time points for \(d = 4\) and \(p = 60\) for \(d = 8\). Panels~(a) and~(c) show the purity of subsystem \(A\) as a function of time, and panels~(b) and~(d) display the corresponding Fourier-mode amplitudes extracted from the simulated data. These results are obtained from numerical simulations only, without hardware execution, since the preparation of spin coherent states requires circuit constructions that become increasingly demanding as \( d \) grows. The simulated curves provide a reference for comparison with implementations that use more straightforward initial states, and, for the small dimensions considered here, we observe that the number of time partitions \( p \) required for Fourier integration can be reduced relative to those implementations, while preserving the algorithmic principles of PIED.

    \end{minipage}

    \label{fig:spin_state_sims}
\end{figure*}

Once the initial states are prepared, the PIED dynamics follow identically to those described in the main text: the bipartite system evolves under the Hamiltonian \(\hat{H}_{AB}\), and the reduced purity of subsystem \(A\) is monitored over time. Specifically, the evolved state reads
\begin{align}
|\psi(t)\rangle_{AB} &= \frac{1}{2^{d-1}}\!\!\!\sum_{n_A, n_B}\!\!\!e^{-i\omega n_An_Bt}\sqrt{
  \bigl(\begin{smallmatrix}
    d - 1 \\
    d - n_A 
  \end{smallmatrix}\bigr)
  \bigl(\begin{smallmatrix}
    d - 1 \\
    d - n_B 
  \end{smallmatrix}\bigr)
}|E_{n_A}E_{n_B}\rangle.
\end{align}
From this evolved state we obtain the reduced density matrix of subsystem \(A\), \(\hat{\rho}_A(t) = \mathrm{Tr}_B \bigl[\,|\psi(t)\rangle_{AB} \langle \psi(t)|\,\bigr]\), and compute the corresponding reduced purity, \(\gamma_A (t) = \operatorname{Tr}(\hat{\rho}_A^2(t))\):
\begin{equation}
    \gamma_A(t) = \frac{1}{16^{d-1}}\!\!\!\sum_{j, k, l, m}\!\!
  \bigl(\begin{smallmatrix}
    d - 1 \\
    d - j 
  \end{smallmatrix}\bigr)
  \bigl(\begin{smallmatrix}
    d - 1 \\
    d - k 
  \end{smallmatrix}\bigr)
  \bigl(\begin{smallmatrix}
    d - 1 \\
    d - l 
  \end{smallmatrix}\bigr)
  \bigl(\begin{smallmatrix}
    d - 1 \\
    d - m 
  \end{smallmatrix}\bigr)e^{-i\omega t(j-k)(l-m)}.
\end{equation}
This is the same quantity accessed in the demonstrations discussed in the main text, but now evaluated for an initial state that is not the uniform superposition over the computational basis, and therefore does not distribute amplitude homogeneously over the energy eigenstates. We extract the Fourier-mode amplitudes \(\alpha_n\) as follows: 
\begin{align}
    \alpha_n &= \left(\frac{4}{16^{d-1}}\right)\sum_{k,m}\sum_{j > k}\sum_{l > m}
  \bigl(\begin{smallmatrix}
    d - 1 \\
    d - j 
  \end{smallmatrix}\bigr)
  \bigl(\begin{smallmatrix}
    d - 1 \\
    d - k 
  \end{smallmatrix}\bigr)
  \bigl(\begin{smallmatrix}
    d - 1 \\
    d - l 
  \end{smallmatrix}\bigr)
  \bigl(\begin{smallmatrix}
    d - 1 \\
    d - m 
  \end{smallmatrix}\bigr) \notag \\
  &\quad\times \delta^{n}_{(j-k)(l-m)}.
\end{align}
The lower bound \(B_n\), which corresponds to prime \(n\) values and respects \(B_n \leq \alpha_n\), is given by: 
\begin{align}
B_n &= 8 \sum_{k = 1}^{d - n} \sum_{m = 1}^{d - 1} |c_k|^2 |c_m|^2 |c_{k + n}|^2 |c_{m + 1}|^2 \notag \\
\label{Bn_spin_coh}
&= \left(\frac{8}{16^{d-1}}\right)\sum_{k = 1}^{d - n} \sum_{m = 1}^{d - 1}
  \bigl(\begin{smallmatrix}
    d - 1 \\
    d - k
  \end{smallmatrix}\bigr)
  \bigl(\begin{smallmatrix}
    d - 1 \\
    d - m
  \end{smallmatrix}\bigr)
  \bigl(\begin{smallmatrix}
    d - 1 \\
    d - k - n
  \end{smallmatrix}\bigr)
  \bigl(\begin{smallmatrix}
    d - 1 \\
    d - m - 1
  \end{smallmatrix}\bigr).
\end{align}
In the simulations presented here, we used \(2^{13}\) measurement shots per time point and chose \(p = 30\) for \(d = 4\) and \(p = 60\) for \(d = 8\), as indicated in Fig.~\ref{fig:spin_state_sims}. No error-mitigation techniques, such as the CFE procedure discussed in the main text, were applied here; all results were obtained directly from noiseless numerical simulations.

It is important to emphasize that the interpretation of the Fourier modes \(\alpha_n\) remains consistent with the prime/composite identification logic established in Sec.~\ref{sec:theoretical}. In particular, the lower bounds \(B_n\) associated with the trivial divisors of \(n\) continue to serve as reference thresholds for distinguishing prime-number signatures. However, since the spin coherent state in Eq.~(\ref{bip_spin_state}) does not uniformly populate the computational basis, the resulting Fourier spectrum acquires a distinct envelope that reflects the nonuniform amplitude distribution of the initial state.

The purity curves and corresponding Fourier spectra shown in Fig.~\ref{fig:spin_state_sims} illustrate these effects. Although their profiles differ from those obtained for uniform initial states, the same characteristic modulation pattern appears in the Fourier components, demonstrating that PIED remains sensitive to the arithmetic structure of \(n\) even when initialized with nonuniform, physically motivated states such as spin coherent ones. These results highlight the generality of the algorithm’s spectral response and its robustness to different initial conditions.

%





\end{document}